\documentclass[journal, onecolumn,12pt]{IEEEtran}

\usepackage{setspace}
\usepackage[cmex10]{amsmath}
\usepackage{amssymb}
\usepackage{amsthm}
\usepackage{array}
\usepackage{graphicx}
\usepackage{dsfont}
\usepackage{array}
\usepackage{cite}
\usepackage{bm}
\usepackage{multirow}	
\usepackage{algorithm}
\usepackage{algpseudocode}

\newcommand\T{\rule{0pt}{2.6ex}}
\newcommand\B{\rule[-1.2ex]{0pt}{0pt}}

\usepackage{tikz}

\newtheoremstyle{example}{\topsep}{\topsep}{}{}{\itshape}{:}{.5em}{\thmname{#1}\thmnumber{ #2}\thmnote{ (#3&)}}
\newtheoremstyle{examplecontd}{\topsep}{\topsep}{}{}{\itshape}{:}{.5em}{\thmname{#1}\thmnumber{ #2}\thmnote{ #3&}\enspace(Cont'd)}
\theoremstyle{example}
\newtheorem{example}{Example}

\theoremstyle{example}
\newtheorem{theorem}{Theorem}
\newtheorem{lemma}{Lemma}

\newtheorem{remark}{Remark}

\def\remark{
  \let\go\relax
  \ifvmode\vskip-\lastskip\fi
  \noindent{\it Remark\/.}%
  \enskip\relax\ignorespaces\go}
\newcommand{\bs}[1]{\ensuremath{\boldsymbol{#1}}}

\begin{document}

\title{
{ML and Near-ML Decoding of LDPC Codes Over the BEC: Bounds and Decoding Algorithms}}

\author{Irina~E.~Bocharova,~\IEEEmembership{Senior~Member,~IEEE},
             Boris~D.~Kudryashov,~\IEEEmembership{Senior~Member,~IEEE},
Vitaly~Skachek,~\IEEEmembership{
Member,~IEEE},
Eirik~Rosnes,~\IEEEmembership{Senior~Member,~IEEE}, \\
and {\O}yvind~Ytrehus,~\IEEEmembership{Senior~Member,~IEEE}
            \thanks{
I.~Bocharova and B.~Kudryashov are with Dept. of Information Systems,
{St.-Petersburg} University of Information Technologies, 
Mechanics and Optics, {St.-Petersburg}, 197101, Russia 
(e-mail: \{iebocharova, bdkudryashov\}@corp.ifmo.ru).

I.~Bocharova and V.~Skachek are with Institute of Computer Science, University of Tartu 
(e-mail: \{irinaboc, vitaly\}@ut.ee).

E.~Rosnes is with Simula@UiB, and {\O}.~Ytrehus is  with Simula@UiB 
and with Dept. of Informatics, University of Bergen
(e-mail: \{eirik, oyvind\}@ii.uib.no).

This paper was presented in part at the 
9th International Symposium on Turbo Codes and Iterative Information Processing, 
Brest, France, September 2016~\cite{bocharova2016wrap}, and 
at the
IEEE International Symposium 
on Information Theory, Aachen, Germany, June 2017~\cite{bocharova2017FL}.

{This work is supported in part by the Norwegian-Estonian Research Cooperation Programme under the grant EMP133, and by the Estonian Research Council under the grants PUT405 and PRG49.} 

Copyright \copyright 2017 IEEE
}
}

%

\maketitle

\begin{abstract}
The performance of the maximum-likelihood (ML) decoding on the binary erasure channel for finite-length low-density parity-check (LDPC) codes from two random ensembles is studied. The theoretical  average spectrum of the Gallager ensemble is computed by using a recurrent procedure and  compared to the empirically found  average spectrum for the same ensemble as well as to the empirical average spectrum of the Richardson-Urbanke  ensemble and spectra of selected codes from both ensembles. Distance  properties of the random codes from the Gallager ensemble are discussed. A tightened  union-type upper bound on the ML decoding error probability based on the precise coefficients of the average spectrum is presented. A new upper bound on the ML decoding performance of LDPC codes from the Gallager ensemble based on computing the rank of submatrices of the code parity-check matrix is derived. A new lower bound on the ML decoding threshold followed from the latter error probability bound is obtained. An improved lower bound on the error probability for codes with known estimate on the minimum distance is presented as well.  A new low-complexity near-ML decoding algorithm for quasi-cyclic LDPC codes is proposed and simulated. Its performance is compared to  the simulated BP decoding performance  and  simulated performance of the best known improved iterative decoding techniques, as well as, with  the upper bounds on the ML decoding performance and decoding thresholds obtained by the density evolution technique.   
\end{abstract}
\section{Introduction}
A binary erasure channel (BEC) is one of the simplest to analyze  and consequently a well-studied communication channel model. In spite of its simplicity, during the last decades the BEC  has started to play a  more important 
role due to the emergence of new applications.
For example,
in communication networks virtually all errors 
occurring 
at the physical level can be detected 
using a rather small redundancy. 
Data packets with detected errors can be viewed as symbol erasures.


{ The analysis of 
the decoding 
 performance on the BEC 
is simpler than for 
{other} channel models.}
{On the other hand, it is expected that ideas and findings for the BEC}
might be useful for constructing 
codes and 
{developing} decoding algorithms for other 
{important 
communication channels}, such as, for example, the binary symmetric channel (BSC) or the additive white Gaussian noise (AWGN) channel.

A commonly used approach to the analysis of decoding algorithms is to study the performance of the algorithm applied to random linear codes over a given channel model. Two of the most studied code  ensembles are the classical Gallager ensemble \cite{gallager} and the more general ensemble presented in  \cite{richardson2008modern}.
The Gallager ensemble  is historically the first thoroughly studied  
ensemble of regular LDPC codes. 
The ensemble in  \cite{richardson2008modern}
can be described by random Tanner graphs with given parity-check and symbol node    
degree distributions. We will refer to this ensemble and the LDPC codes contained in it as the  {\it Richardson-Urbanke} (RU) ensemble and RU LDPC codes, respectively. 
These and some  other ensembles of regular LDPC codes are described and analyzed in \cite{litsyn2002ensembles}.

It is shown in \cite[Appendix~B] {gallager}  that  the asymptotic 
weight enumerator  of  random $(J,K)$-regular LDPC codes approaches the asymptotic weight enumerator of random linear codes when $J$ and $K$ grow.
In \cite{litsyn2002ensembles}, it is confirmed that 
other ensembles of regular LDPC codes demonstrate  a similar behavior.    
Thus, regular ensembles are good enough to achieve near-optimal performance.  On the other hand, it is well-known that both asymptotically \cite{richardson2008modern} and in a finite-length regime  
irregular 
LDPC codes outperform their regular counterparts and more often are recommended   
for real-life applications \cite{STD802,STDDVB}.
Finite-length  analysis of RU LDPC codes under belief propagation (BP) and (to a lesser degree) under maximum-likelihood (ML) decoding was performed in 
\cite{di2002finite}. In this paper, we extend the analysis  of the ML decoding case for regular codes. In particular, new error probability bounds 
for regular codes are presented. For general linear codes, detailed overviews of lower and upper bounds 
for the AWGN channel and the BSC  were 
presented by 
Sason and Shamai  in their tutorial paper \cite{sason2006performance}, and for the AWGN channel, the BSC, and the BEC by 
Polyanskiy et al.\ in \cite{polyanskiy2010channel} and by Di et al.\ in \cite{di2002finite}. 

For computing upper bounds on the error probability for LDPC codes  there exist two approaches. One, more general,  approach is based on a union-type 
bound and requires  knowledge of the code weight enumerator or its estimate.  The second approach used in \cite{di2002finite} is suitable for the BEC. It  implies estimating the rank of submatrices of the LDPC code parity-check matrix. Notice that  for infinitely long codes,  the bit error rate (BER) performance of BP decoding can be analyzed through density evolution (see e.g. 
\cite{luby2001efficient,richardson2008modern}). However, the density evolution technique is not
suitable for analysis of finite-length codes, since 
dependencies caused by cycles in the Tanner graph associated with the code
are not taken into account. 

In this paper,  first we consider a tightened union-type bound based on precise average spectra of random 
finite-length LDPC codes.   
The difference between our approach and other techniques is the way of computing  the 
bound.  Instead of manipulating with hardly computable coefficients of series
expansions of generating functions, we 
compute the spectra by using efficient recurrent procedures. This allows for obtaining 
precise average weight enumerators with complexity growing linearly with the code length.   
New bounds, based on computing the rank of submatrices, are  derived for the RU and the Gallager ensemble of regular LDPC codes.
A tightened lower bound on the ML  decoding error probability on the BEC, which is 
applicable to any linear code with a known upper bound on the minimum distance, is presented. For LDPC codes of short length, it shows a surprisingly strong improvement  over the best known lower bound.   A lower  bound on the ML decoding threshold which stems from the derived upper rank-type bounds  is derived. Notice that this bound  obtained from the bound on the average block  error probability over the Gallager and RU  ensembles slightly  differs  from the ML decoding thresholds   in \cite {sason2003parity} and \cite{measson2008maxwell }  obtained from  the bounds on  the bit error probability of the average code in the RU ensemble.   

{A} remarkable property of 
{the}
 BEC is that ML decoding 
 {of} any linear code 
over this channel is reduced to solving a system of linear equations. 
{This} means that ML decoding 
of an $[n,k]$ LDPC code (where $n$ is the code length and $k$ is the number of information symbols) with $\nu$ erasures
can be performed by Gaussian elimination with time complexity 
at most $O(\nu^3)$. 
Exploiting the sparsity of the parity-check matrix of the LDPC codes can lower the complexity  to approximately $O(\nu^2)$ (see overview and analysis in \cite{burshtein2004efficient} and references therein).
Practically feasible algorithms with a thorough complexity analysis can be found in
\cite{paolini2012maximum, cunche2010analysis, kim2008efficient}. 
This makes ML-decoded LDPC codes  strong 
competitors for scenarios with 
strong restrictions on 
the decoding delay. 
It is worth noting that ML decoding allows for achieving low error probability at 
rates strictly  above the so-called BP decoding threshold \cite{richardson2008modern}. However, ML decoding of long LDPC codes of lengths of tens of  thousands of  bits  over the BEC is still considered impractical unless the code rate $R$ tends to $1$ or number of erasures is small  (see, for example,\cite{ kim2008efficient}).

Low-complexity suboptimal decoding techniques  for LDPC codes over the BEC are based on the following two approaches.  The first approach consists of adding redundant rows to the original code parity-check matrix  (see, for example, 
 \cite{vasic2005,kobayashi2006,Vihula2017}). The second approach applies post-processing in case of BP decoding failure \cite{pishro2004decoding, hosoya2004, olmos2010tree,vellambi2007}.  In \cite{pishro2004decoding,vellambi2007}, a post-processing step based 
on the concept of guessing bits is proposed.
Applying 
bit guessing based algorithms to the decoding of LDPC codes improves the performance of BP decoding, but does not provide near-ML performance. 

In this paper, we propose a new  decoding algorithm which provides near-ML decoding of long quasi-cyclic (QC) LDPC  block codes. 
It is well-known that QC LDPC codes  are  considered
practical in terms of implementation complexity and they are often referred to as candidates for different  communication standards. 
Furthermore, QC LDPC block codes can be represented as a tail-biting (TB) parent convolutional LDPC code. 
Thereby, decoding techniques developed for TB codes are applicable to QC  LDPC codes. 
The proposed algorithm resembles a  wrap-around suboptimal decoding of TB  convolutional codes
\cite{kudryashov1990decoding, shao2003two}.  Decoding of a TB code requires identification of the correct starting state,
and thus 
{ML decoding must apply the Viterbi algorithm  once for} each possible starting state. 
In contrast, wrap-around decoding applies the Viterbi algorithm once to the \emph{wrapped-around}  trellis diagram  with all starting state metrics initialized to zero.  This decoding approach yields near-ML performance at a typical complexity of a few times the complexity
of the Viterbi algorithm. 

The new algorithm belongs to a family of so-called hybrid decoding algorithms. It  is based  on a combination of BP decoding of the QC LDPC code followed by  so-called ``quasi-cyclic sliding-window'' ML  decoding. The latter technique is applied ``quasi-cyclically'' to a
relatively short sliding window, where the decoder performs ML decoding  of a zero-tail terminated  (ZT) LDPC  convolutional code. Notice that unlike  sliding-window near-ML (SWML) decoding of convolutional codes considered in \cite{ tomas2012}, the suggested algorithm working on the parent LDPC convolutional code has significantly lower computational complexity due to the sparsity of the code parity-check matrix \cite{bocharova2016bec}. On the other hand, it preserves almost all advantages of the convolutional structure in the sense of erasure correcting capability.

%

The rest  of the paper is organized as follows. Preliminaries are given in Section~\ref{sec_prilim}.  A recurrent algorithm for computing the average spectrum for the Gallager ensemble of binary LDPC codes is presented in Section~\ref{sec_spec}. Empirical average 
{spectra} for the Gallager and RU ensembles are computed and compared to the theoretical  average  spectra as well as 
{to the} spectra of selected codes in the same section. 
Distance properties of the Gallager ensemble are discussed in 
{Appendix~A}.  In Section~\ref{sec_bound}, an improved lower bound  and two  
{upper bounds} on the error probability of ML decoding over the BEC are derived. The corresponding proofs are given in Appendices~B and C  for the RU and the 
Gallager ensemble, respectively. 
A new algorithm for near-ML decoding of long QC LDPC codes based on the interpretation of these codes as TB convolutional codes and using wrap-around  sliding-window decoding is proposed in Section~\ref{sec_method}. Simulation results confirm  the efficiency of the algorithm and are presented in Section~\ref{numres}, while  
asymptotic  decoding thresholds  which stem from the derived rank bounds are computed in Section~\ref{Sec7}.  Conclusions are drawn  in Section~\ref{sec:conclu}.
 
%

\section{Preliminaries}
\label{sec_prilim}



A binary
linear $[n,k]$ code $\mathcal C$ of rate $R=k/n$ can be defined as the null space of a $r \times n$ binary parity-check matrix $H$ of rank $\rho=n-k \leq r$.
%
%
Denote by  $\{A_{n,w}\}$, $w=0,1,\dotsc,n$,  the code weight enumerator, where $A_{n,w}$ is the number of codewords of 
 weight $w$. 
By some abuse of notation, we use $A_{n,w}$ for both the weight enumerator of a specific code and for 
random weight enumerators in code ensembles.

Random parity-check matrices $H$ of size $r\times n$
{do not} necessarily have rank $\rho=r$ which means that in general $k \geq n-r$. 
%
Following a commonly accepted assumption we assume that $\rho=r$ when deriving bounds on the error probability, but we take the difference into account when considering code examples.

Two ensembles of random regular LDPC codes are studied below.
The first 
{ensemble} is the Gallager ensemble  \cite{gallager} of $(J,K)$-regular LDPC codes.
Codes of this ensemble are determined by  random parity-check matrices $H$  which consist  
of strips  $H_{i}$  of width $M=r/J$ rows each,  $ i =1,2,\dots, J$. 
All strips are random column permutations of the strip  where the $j$-th row 
contains $K$ ones in positions $(j-1)K+1, (j-1)K+2, \ldots, jK$, for $j = 1, 2, \ldots, n/K$.  The second ensemble 
is a special case of the ensemble described in \cite[Definition 3.15] {richardson2008modern}.
Notice that the Gallager ensemble and the RU ensemble are denoted by $\mathcal B$ and $\mathcal H$, respectively, in \cite{litsyn2002ensembles}. 

For  $a\in\{1,2,\dotsc\}$ denote by $a^m$ the sequence $(a,a,\dotsc,a)$  of $m$ identical 
symbols $a$. 
In order to construct an $r \times n$ parity-check matrix $H$ of  an LDPC code from the  RU ensemble perform the following steps.
\begin{itemize}
\item 
Construct the sequence $\bs a=(1^J,2^J,\dotsc,n^J)$; and
\item
apply a random permutation $\pi(\cdot)$ to obtain a sequence $\bs b= \pi (\bs a) =(b_1,\dotsc,b_N)$, where $N=Kr=Jn$. The elements $b_1,\dotsc,b_K$ show the locations  of the nonzero elements of the first row of  $H$, 
elements $b_{K+1},\dotsc,b_{2K}$ show the locations  of the nonzero elements of the second row of  $H$, etc.
\end{itemize}
 
A code from the RU ensemble is $(J,K)$-regular if for a given permutation $\pi$ all elements of subsequences 
$(b_{iK-K+1},\dotsc,b_{iK})$ are different for all $i=1,\dotsc,r$, otherwise it is irregular.  The regular RU codes belong to the ensemble $\mathcal A$ in \cite{litsyn2002ensembles} which is defined as the sub-ensemble of the RU ensemble containing the parity-check matrices with row weight $K$ and column weight $J$. It is shown in \cite{litsyn2002ensembles} that the three ensembles  $\mathcal A$, $\mathcal B$, and $\mathcal H$ have the same asymptotic average weight enumerators.

It is known (see \cite[Theorem~3]{litsyn2002ensembles})  
that for large $n$ the total number of $(J,K)$-regular $[n,n-r]$ codes 
(ensemble $\mathcal A$ in \cite{litsyn2002ensembles}) is 
equal to 
\[
\frac {(Jn)! }{(K!)^r(J!)^n }\exp \left\{-\frac{(K-1)(J-1)}{2}   \right\}\left(1+o(n^{-1+\delta}) \right) \;,
\]
where $\delta>0$ and $o (x) \to 0$ when $x \to 0$.
The number of different codes  from the RU ensemble constructed as described above  is 
\[
\frac {(Jn)! }{(K!)^r(J!)^n } \;.
\]
Thus, the portion of $(J,K)$-regular  LDPC  codes
in the RU ensemble is
\[
\exp \left\{-\frac{(K-1)(J-1)}{2}   \right\}\left(1+o(n^{-1+\delta}) \right) \;,
\]
 that is,  most of the ``$(J,K)$-regular'' RU codes are indeed irregular. However, for a particular parity-check matrix, the fraction of  rows  of weight less than $K$ as well as  the fraction of columns of weight less than $J$ is small (both fractions tend to zero when $n$ grows). Thus, this irregularity can be safely ignored. 
 {In the following, a code from the RU ensemble with parameters $J$ and $K$ will sometimes be referred to as a $(J,K)$-RU code or simply as a $(J,K)$-regular code even if it is not strictly $(J,K)$-regular. Also, with some abuse of language a $(J,K)$-regular code from the Gallager ensemble will be referred to as a $(J,K)$-Gallager code.}

As a performance measure we use the word (block, frame) error rate (FER) $P_e$,
which for the BEC is defined as the probability that the decoder cannot
recover the information of a received word uniquely.

Consider ML decoding over the BEC, where 
$\varepsilon>0$ denotes the channel symbol erasure probability.
Assume that a  codeword $\bs x=(x_1,\dotsc,x_n)$ is transmitted and that $\nu$ erasures occurred. 
Then, we denote by $I$  the set of  
{indices} of the erased positions, that is, 
 $I=\{i_1,\dotsc, i_{\nu}\}\subseteq \{1,2,\dotsc,n\}$, $|I|=\nu$, and by
 $\bm x_I=(x_{i_1}, \dotsc, x_{i_{\nu}})$ a vector of unknowns corresponding to the erased positions. Let  $I^{\rm c}=\{1,2,\dotsc,n \}\setminus I$ and  $\bm x_{I^{\rm c}}$  be the set of 
 {indices} 
 of unerased positions and the vector of unerased values of $\bs x$, respectively.   Also, let $H_I$  be the submatrix of $H$ consisting of the columns indexed by $I$.
From $\bs x H^{\rm T}=\bs 0$, where $(\cdot)^{\rm T}$ denotes the transpose of its argument, it follows that
\begin{equation}
\bs x_I H_I^{\rm T} = \bs x_{I^{\rm c}} H_{I^{\rm c}}^{\rm T}\triangleq {\bs s} \;,
\label{eq_syndrome}
\end{equation}
where $\bs s$ is the syndrome vector
or, equivalently,
\begin{equation} \label{main1}
\bs y_I   H_I^{\rm T}=\bs 0 \;,
\end{equation}  
where $\bs y=\bs x +\bs x'$ is a codeword.
If the solution of (\ref{main1}) is not unique, that is, 
\begin{equation*} 
\mbox{rank}\left( H_{I}\right)<|I| \;,
\end{equation*}
where $\mbox{rank}\left(\cdot\right)$ denotes the matrix rank of its argument, 
then the corresponding set of erasures cannot be (uniquely) corrected. 
Otherwise, the set of erasures $I$ is correctable. Thus, 
{the
ML decoding error probability (for the BEC) is the probability of such a set $I$, that is,} 
\begin{equation} \label{main3}
P_e=\Pr \left(\mbox{rank}\left( H_{I}\right)<|I| \right) \;.
\end{equation}

\section{Average spectra for ensembles of regular LDPC codes \label{sec_spec}}

{
\subsection{Weight enumerator generating functions}
}

In this section, we study average weight enumerators for  different ensembles of LDPC codes.
The  weight  distribution  of any  linear code  can be represented in terms of its weight generating function 
\[{G_{n}(s)=\sum_{w=0}^{n}A_{n,w}s^{w} \; ,} \]
where $A_{n,w}$ is the random variable representing the number of binary codewords of weight $w$  and length $n$, 
{and $s$ is a formal variable}. 
Our goal is to find ${\rm E}\{A_{n,w}\}$, where $\rm E\{\cdot\}$ denotes the expected value over the code ensemble. In general,  computing the coefficients ${\rm E}\{A_{n,w}\}$  is a rather difficult task. 
If a generating function  can be represented as a degree of another generating function 
(or expressed via a degree of such a function),  then 
for numerical  computations we can  use the following simple 
recursion.

{\begin{lemma}
\label{Lemma}  
Let  $f(s)=\sum_{l \ge 0} f_{l}s^{l}$ be a generating function.  
Then,  the coefficients in the series expansion of the  generating function $F_{L}(s)=\left [f(s)\right]^{L}=\sum_{l\ge 0}F_{l,L}s^{l}$  satisfy the following  recurrent equation
\begin{eqnarray*}
F_{l,L}=\left\{ \begin{array}{ll} 
    f_{l},  &L=1 \\
\sum_{i=0}^{l}f_{i}F_{l-i,L-1}, &L>1 
\end{array} \right. \; . 
\end{eqnarray*}
\end{lemma}

\subsection {General linear codes \label{rec_sp}}
For completeness, we 
{present} the average spectrum for the ensemble of  random linear codes determined by  equiprobable  $r\times n$ parity-check matrices, where $r=n-k$, and $k$  and $n$ are the code dimension and length, respectively.  The weight generating function of all binary sequences of length $n$ is   $G_{n}(s)=(1+s)^{n}$. Then, the average  spectrum coefficients  are
\begin{equation} 
{\rm E} \{A_{n,w} \}=\binom{n}{w}2^{-r}\;, \; w>0 \;,  
\label{gen}
\end{equation} where $2^{-r}$ is the probability that a binary  sequence $\bs x$ of length $n$ and weight $w>0$ satisfies $ \bs x H^{\rm T}=\bs 0 $.  

If a random linear code contains only codewords of even weight, then its generating function has the form
\[G_{n}(s)=\sum_{w\; \rm{even}}\binom{n}{w}s^{w}= \frac{(1+s)^{n}+(1-s)^{n}}{2} \;,\]
and the average spectrum coefficients are
\begin{equation}
{\rm E}\{A_{n,w}\}=
\left\{
\begin{array}{ll}
2^{-r+1}\binom{n}{w}, & w>0 \mbox { and even} \\
0, &w                                \mbox{ odd} 
\end{array}
\right. \;.
\label{lin_code}
\end{equation}

\subsection {The Gallager binary $(J,K)$-regular random LDPC codes \label{sec:1.2}}
The generating function of the number of sequences satisfying the nonzero part of one parity check 
is given by
\begin{equation}\label{pdw}
g(s)=\sum_{i\; \rm{even}} \binom{K}{i}s^{i}=\frac{1}{2}\left[(1+s)^K +(1-s)^K \right]  \;.
\end{equation}
The generating function for the strip is
\begin{equation} \label{bin_gen_fun}
G_{J,K}(s) =g(s)^M=\sum_{w=0}^{n} N_{n,w}s^{w} \;, 
\end{equation}
where $N_{n,w}$ denotes the total number of binary sequences of weight  $w$ satisfying
 $\bs x H_1^{\rm T}= \bs 0$. 
Due to Lemma  \ref{Lemma} we can compute  $N_{n,w}$  precisely.
The probability that  $\bs x H_1^{\rm T}= \bs 0$  
is valid for a random 
{binary} 
$\bs x$ of weight $w$ is equal to
\[
p_1(w)=\frac{N_{n,w}}{\binom{n}{w}} \;.
\]

Since the submatrices  $H_j$, $j=1,\dotsc,J$, are obtained by independent random column permutations of $H_1$,
the expected number of codewords among all $\binom{n}{w}$ sequences of weight $w$ is  
 \begin{equation} \label{bin_spec}
{\rm E}\{A_{n,w}\}=\binom{n}{w}   p_1(w)^J =
   \binom{n}{w}^{1-J}  N_{n,w}^J \;,
\end{equation} 
where $N_{n,w}$ is computed recursively using Lemma \ref{Lemma}.

\subsection{The empirical and theoretical average spectra for the Gallager and the Richardson-Urbanke ensemble}
 \label{Gal_app}  

In this section, we compare the theoretical average spectra computed according to  
(\ref{gen}), (\ref{lin_code}), and (\ref{bin_spec}) with the empirically obtained average spectra for the Gallager and the RU ensemble. Furthermore, we compare the average spectra with the spectra of both randomly generated and constructed LDPC and linear block codes.  

{We remark that} there is a weakness in the proof of Theorem~2.4 in \cite{gallager} by Gallager, 
{which is similar to the one in the}
derivations (\ref{pdw})--(\ref{bin_spec}) above.  Formula (2.17)  in
\cite{gallager} (and (\ref{bin_spec}) in this paper) states that
the average number of 
weight-$w$ binary sequences  which satisfy the parity checks of  all $J$  strips simultaneously   
is obtained by computing  the $J$-th degree of  $N_{n,w}/\binom{n}{w}$, that is, the probability of weight-$w$ 
{sequences} satisfying 
{the parity checks of the first strip.}  
This formula relies  on the assumption that parity checks of strips are statistically independent. 
{Strictly speaking, this} statement is not true because they are always linearly dependent (the sum of the parity checks of any two strips is equal to the all-zero sequence). The detailed discussion and examples can be found in Appendix~A.

In our derivations of the  bounds on the error probability in Section~\ref{sec_bound} we rely on the same assumption. Hence,  it is important to compare the empirical and the theoretical average spectra. Moreover, 
 as it is shown in Appendix~A,  in the Gallager ensemble there is no code whose spectrum coincides with the average spectrum. Thus,  estimating the deviation  of the spectrum of a particular code from the average spectrum is an interesting issue.  One more question 
{that we try to answer} in this section is how close are 
{the} average spectra for the Gallager and RU ensembles. It is known \cite{litsyn2002ensembles} that  the Gallager and RU ensembles have the same asymptotic average spectrum. However, 
{their relation for finite lengths is unknown.}     



{
In Figs.~\ref{g36}--\ref{ru48}, the distance spectra of $100$ randomly selected rate $R=\frac{1}{2}$ 
codes of length $n=48$ (``Random codes'' in the legends) and their empirical average spectrum (``Empirical average'' in the legends)
are compared with the theoretical average spectra.}
We take into account that all codewords of a  $(J,K)$-regular LDPC code  from the Gallager ensemble have even weight.
If $K$ is even, then the all-one codeword belongs to any code from the ensemble.   
It is easy to see that in this case the weight enumerators $A_{n,w}$  
are symmetric functions of $w$, that is, $A_{n,w}=A_{n,n-w}$, and hence we show 
only half of the spectra in these figures. 

\begin{figure}
\begin{center}
\includegraphics[width=85mm]{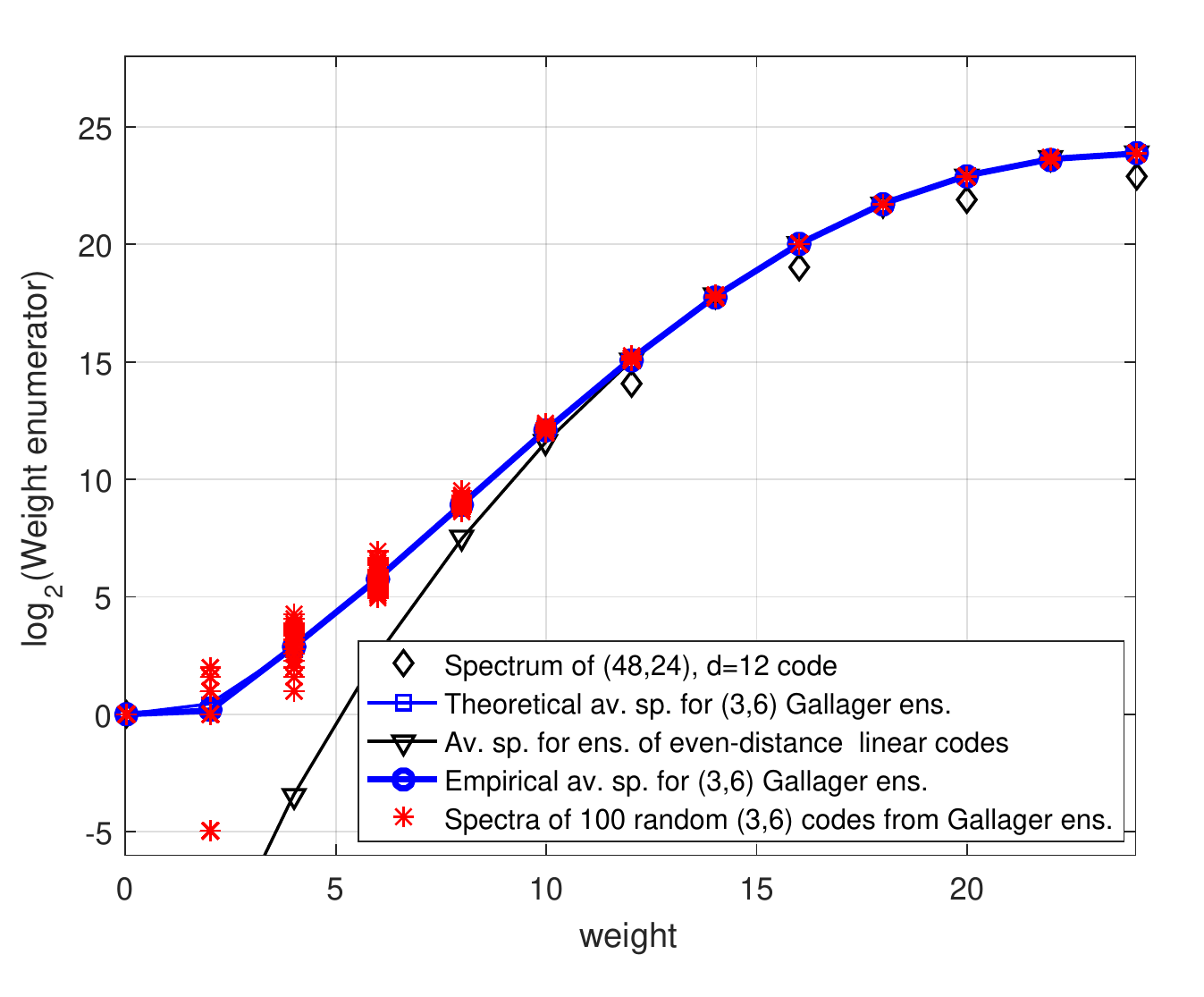}   
\caption{\label{g36} The theoretical and empirical spectra of $(3,6)$-regular Gallager codes of length  $n=48$. 
The average spectrum for the Gallager ensemble and for the ensemble of even-distance linear codes  are defined by (\ref{bin_spec})
 and (\ref{lin_code}), respectively.
}

\end{center}
\end{figure}

\begin{figure}
\begin{center}
\includegraphics[width=85mm]{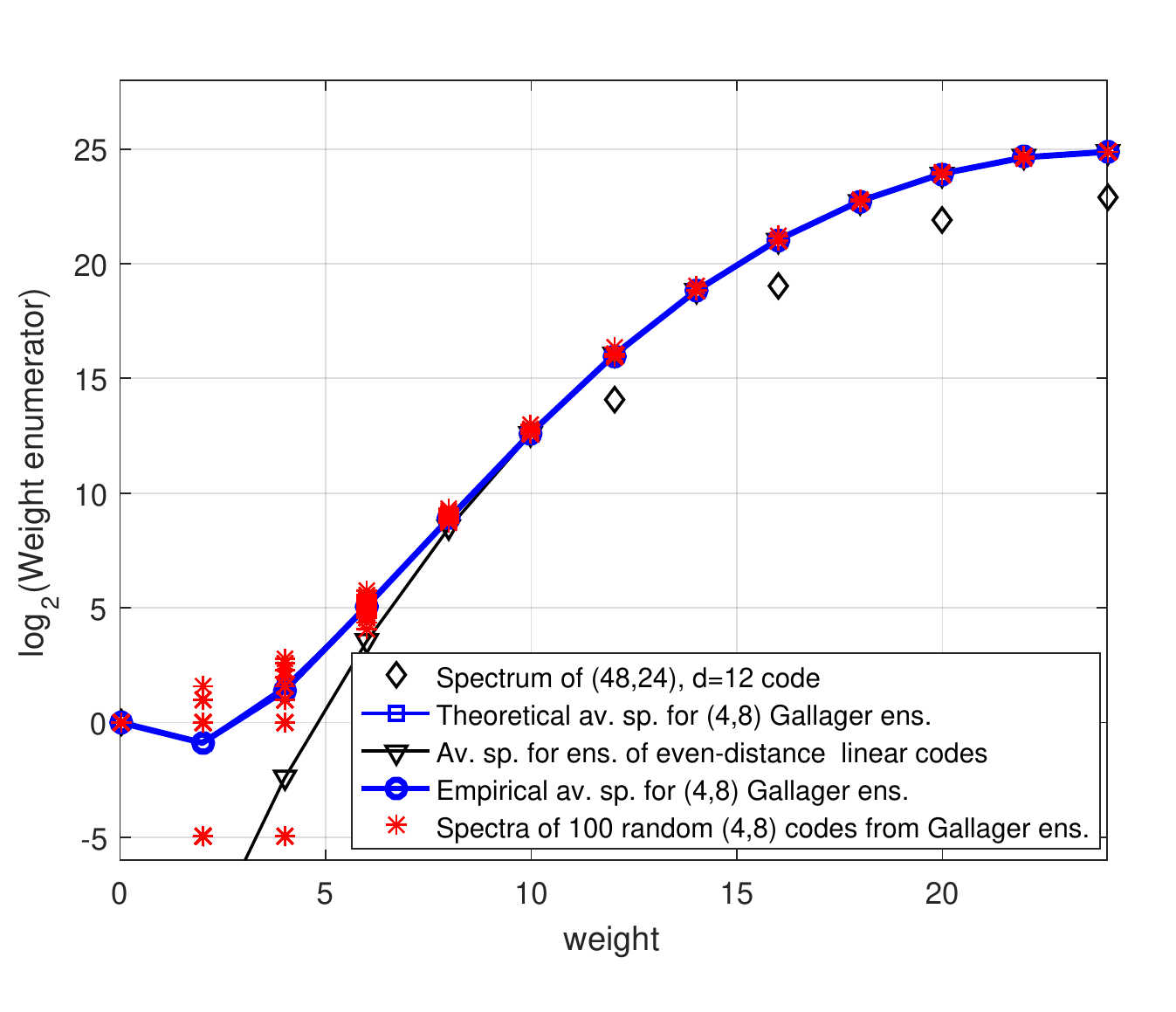}   
\caption{\label{g48}  The theoretical and empirical  spectra of $(4,8)$-regular Gallager codes of length  $n=48$.
The average spectrum for the Gallager ensemble and for the ensemble of even-distance linear codes  are defined by (\ref{bin_spec}) and (\ref{lin_code}), respectively.
  }
\end{center}
\end{figure}
\begin{figure}

\begin{center}
\includegraphics[width=85mm]{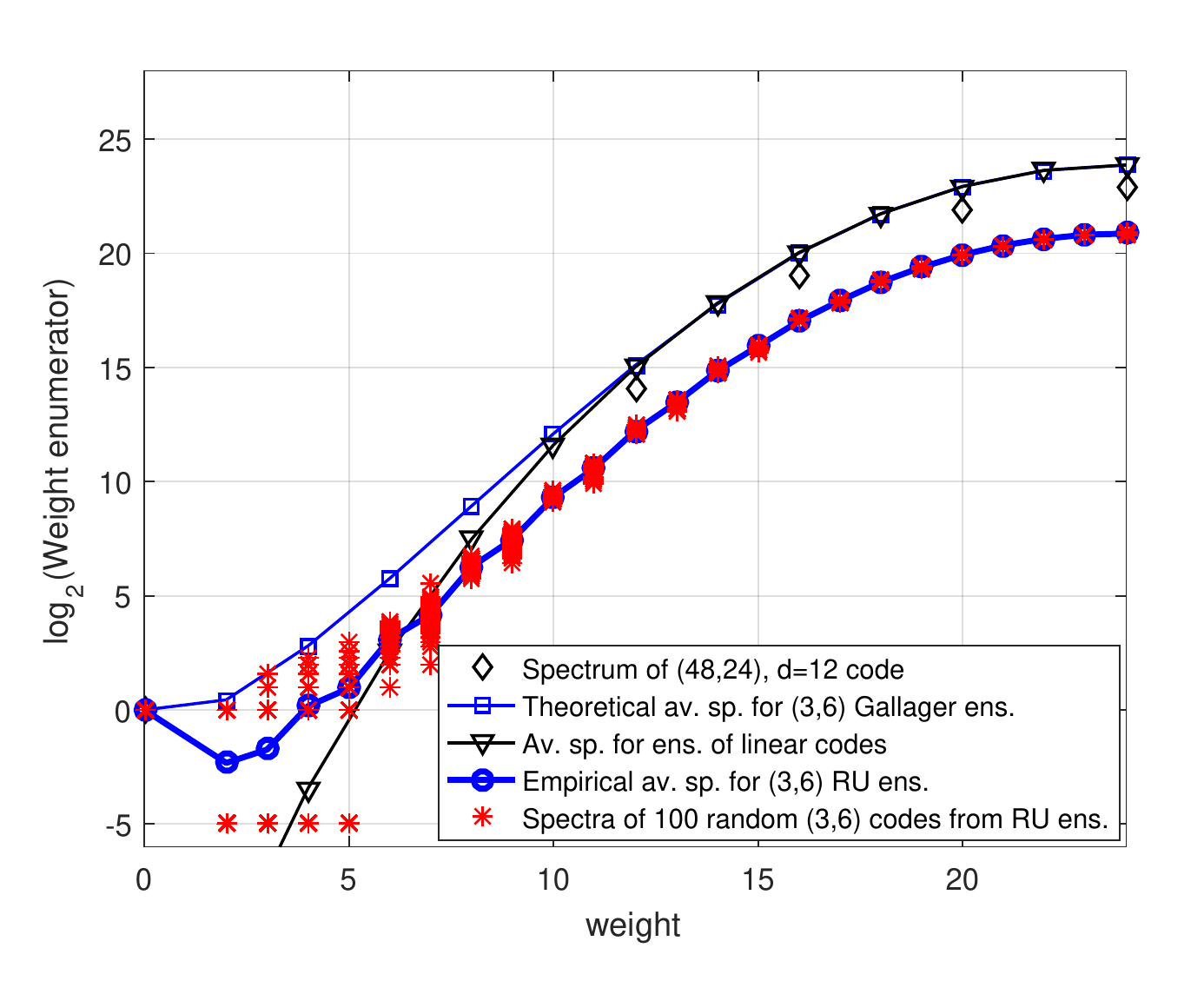}   
\caption{\label{ru36}  The theoretical spectrum of   $(3,6)$-regular Gallager   
codes and the empirical spectra of $(3,6)$-RU LDPC codes of length  $n=48$. 
The average spectrum for the Gallager ensemble and for the ensemble of random linear codes  are defined by (\ref{bin_spec}) and (\ref{gen}), respectively.
}
\end{center}
\end{figure}

\begin{figure}
\begin{center}
\includegraphics[width=85mm]{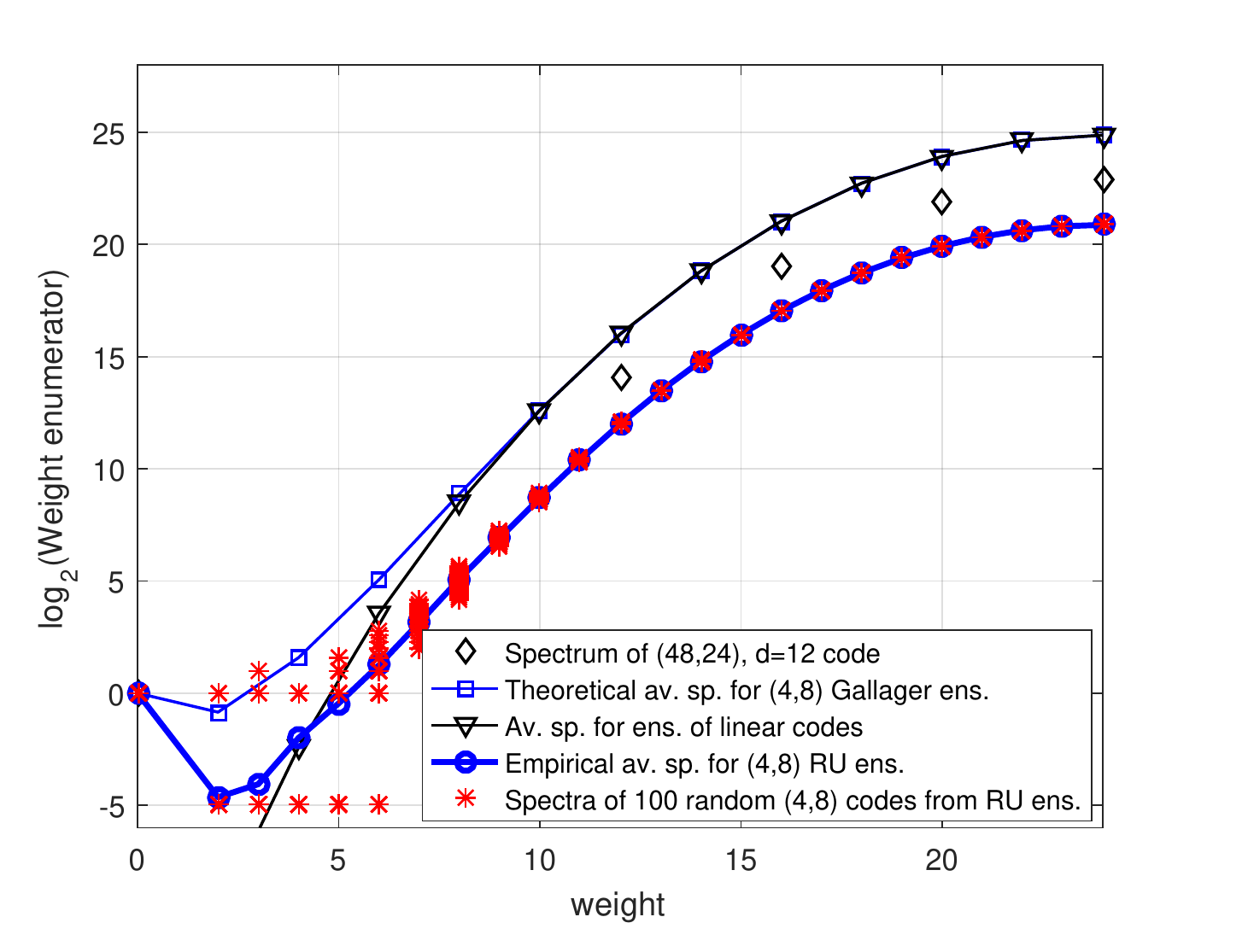}   
\caption{\label{ru48}  The theoretical spectrum of $(4,8)$-regular Gallager  codes and the empirical spectra of 
$(4,8)$-RU LDPC codes of length  $n=48$. 
The average spectrum for the Gallager ensemble and for the ensemble of random  linear codes  are defined by (\ref{bin_spec}) and (\ref{gen}), respectively.
  }
\end{center}
\end{figure}

In Figs.~\ref{g36} and \ref{g48}, we present
 the average spectrum for the  Gallager ensemble and the average
 spectrum for the ensemble of  random linear codes with only even-weight codewords, computed using formulas from Section~\ref{rec_sp},  
spectra of $100$ random codes from the Gallager ensemble, and the empirical  average spectrum  computed
over $100$ random codes from the same ensemble.  The spectrum of a quasi-perfect $[48,24]$ linear code with minimum distance $d=12$ is presented in the same figures as well. 
In Figs.~\ref{ru36} and \ref{ru48}, the average spectrum for the  Gallager ensemble and the average spectrum for the ensemble of random linear codes
are compared with the spectra of $100$ random codes from the RU ensemble and the empirical average spectrum computed over $100$ random codes from the RU ensemble.
In Figs.~\ref{g36}--\ref{ru48}, we use a negative value ($-5$)  for  the logarithm of a spectrum coefficient 
to indicate} that the corresponding coefficient is zero.

We make the following observations regarding the Gallager codes:
\begin{itemize}
\item For the Gallager $(3,6)$ and $(4,8)$-regular LDPC codes their empirical  average spectra perfectly match with the theoretical average spectra computed for the corresponding ensembles.
\item For all codes from the Gallager ensemble the number of high-weight codewords is perfectly predicted by 
the theoretical average spectrum.
\item   The number of low-weight codewords has large variation.
\end{itemize}

We make the following remarks about the RU codes:
\begin{itemize}
\item Most of the RU LDPC codes are indeed irregular and have codewords of both even and odd weight.
\item Typically, parity-check matrices  of random codes from the RU ensemble have full rank, and these codes have  lower  rate than LDPC codes from the  Gallager ensemble.
For this reason, the empirical  average spectrum of the RU ensemble lies below the theoretical average spectrum computed for the Gallager ensemble.
\item The average distance properties of  the  RU codes are better than those  of the  Gallager codes.
\item The variation of the number of low-weight codewords is even larger than that for the Gallager codes.
\end{itemize}

Since for all considered ensembles low-weight codewords have a much larger probability than for 
general linear codes, 
{we expect to observe a higher error floor.}

\section{Error probability bounds  on the BEC \label{sec_bound}}

In this section, we present a new tightened lower bound (Theorem~\ref{lower}) on the ML decoding error probability for  linear codes   and  two new upper bounds (Theorem~\ref{RankLdpc}, Theorem~\ref{RankGal}) on the ML decoding error probability for the RU and Gallager ensembles, respectively.

\subsection{Lower bounds}
We start with a simple lower bound which is true for any linear code.
\begin{theorem} \label{triv}
\begin{equation}
P_e \ge  P_{e}(n,k,\varepsilon)\triangleq \sum_{i=r+1}^n \binom{n}{i} 
\varepsilon^i(1-\varepsilon)^{n-i}\;.
\label{lower_b}
\end{equation}
\end{theorem}

\begin{remark}
\cite[Theorem~38] {polyanskiy2010channel} gives a  lower bound on the error probability 
of ML  decoding. It differs from (\ref{lower_b}) by a multiplier which is close to $1$.
This difference appears because of different definitions of the frame error event in this paper 
and in \cite[Theorem~38] {polyanskiy2010channel}. 
\end{remark} 
   
\begin{IEEEproof}
It readily follows from the definition of the decoding error probability and from the condition in (\ref{main3}) that if 
the number of erasures $\nu>r\ge \mbox{rank}\left( H \right)$, then  the decoding error probability is equal to one.
\end{IEEEproof}

The bound (\ref{lower_b}) ignores erasure combinations of weight 
{less than} or equal to $r$. Such combinations lead to an error if they 
cover all nonzero entries of a codeword.
\begin{theorem}\label{lower}
Let the code minimum distance 
{be $d_{\min}\le d_{0}$. Then,}   
\begin{equation} \label{LT}
P_e \ge  P_{e}(n,k,\varepsilon)+
 \sum_{w=d_0}^{r} \binom{n-d_0}{w-d_0}\varepsilon^w(1-\varepsilon)^{n-w}  \; .
\end{equation}
 \end{theorem}
 \begin{IEEEproof}
There is at least one nonzero codeword $\bs c_0$ of weight at most $d_0$ in the code.
Each  erasure combination of weight $w\ge d_0$ which covers the nonzero positions of $\bs c_0$
leads to additional decoder failures  taken into account as 
{the} sum in 
{the} right-hand side (RHS) of (\ref{LT}).
\end{IEEEproof}
{We remark} that 
upper bounds on the  minimum distance of linear codes with given $n\le 256$ and $k$ can be found in \cite{GrasslCodetables}.    The lower bounds  (\ref{lower_b}) and (\ref{LT})   are compared in Fig. \ref{BEC_bounds4}.

%
%
%

\subsection{Upper bounds for general linear codes}
Next, we consider  the ensemble-average ML decoding block error probability ${\rm E}\{P_{e}\}$
over the BEC with erasure probability $\varepsilon>0$. This average decoding error probability can be interpreted as an upper bound on the achievable error probability for codes from the ensemble. In other words, there exists at least one code in the ensemble whose ML decoding error probability is upperbounded by ${\rm E}\{P_{e}\}$.
To simplify notation, in the sequel we use $P_{e}$ for the ensemble-average error probability. 
For  the ensemble  of random  binary $[n,n-r]$ linear codes 
\begin{eqnarray}
P_{e}
&=&  \sum_{\nu=r+1}^n\binom{n}{\nu}\varepsilon^{\nu}(1-\varepsilon)^{n-\nu} \nonumber\\
&+&\sum_{\nu=1}^{r}\binom{n}{\nu}\varepsilon^{\nu}(1-\varepsilon)^{n-\nu}
P_{e|\nu} \;,  \label{genbound}
\end{eqnarray}
where $P_{e|\nu}$ denotes the conditional ensemble-average error probability given that $\nu$ erasures occurred.

By using the approach  based on estimating the rank of submatrices of random matrices
\cite{landsberg1893ueber},  the expression
 \begin{equation}
P_{e|\nu}=
\Pr\left(\mbox{rank}(H_I) < \nu\right) = 1-
\prod_{j=0}^{\nu-1}\left( 1-2^{j-r}    \right)
 \le  2^{\nu-r} \;,
 \label{precise}
\end{equation}
where  $H_I$ is an $r\times \nu$   submatrix of 
{a random} $r\times n$
parity-check matrix $H$, was obtained  for $P_{e|\nu}$ 
in \cite{berlekamp1980technology,macmullan1998comparison,di2002finite}.

The bound obtained by combining (\ref{genbound}) and (\ref{precise}) is used as a benchmark to compare 
the FER performance of ML decoding of LDPC codes to the FER performance  
of general linear codes in Figs.~\ref{BEC_bounds4}--\ref{WML}.

{An alternative upper bound  for a specific linear code with known weight enumerator  has the form
\cite{berlekamp1980technology}
\begin{equation} \label{gen_spectr}
P_e \le  \sum_{i=d_{\rm min}}^n \min \left\{  \binom{n}{i} ,
\sum_{w=d_{\rm min}}^i A_{n,w} \binom{n-w}{i-w}
 \right\} \varepsilon^i(1-\varepsilon)^{n-i} \;.
\end{equation}
In particular, this bound can be applied to random ensembles of codes with known 
average spectra (see Section \ref{sec_spec}). 
We refer to this bound as the {\it S-bound}. It is presented for several ensembles in 
Figs.~\ref{BEC_bounds4}--\ref{WML}
and discussed in Section~\ref{numres}.

\subsection{Random coding  upper bounds for $(J,K)$-regular LDPC codes }

In this subsection, we derive an upper bound on the ensemble-average error 
probability of ML decoding of the RU and Gallager ensembles of $(J,K)$-regular LDPC 
codes. 
{Similarly to the approach in \cite{di2002finite},} we estimate the rank  of the submatrix $H_{I}$.

A generalization of the bound in (\ref{precise}) to an ensemble of
LDPC codes over the $q$-ary BEC is presented in \cite{liva2013bounds}. Notice
that the upper bound in \cite{liva2013bounds} for $q$ = 2 is derived for a sparse
parity-check code ensemble with column and row weights
growing with the code length.

\begin{theorem}\label{RankLdpc} The $(J,K)$-RU ensemble-average ML decoding error probability  for 
{$[n,n-r]$ codes,} $n=MK$, $r=MJ$,  and $M \gg 1$, is 
{upperbounded} by
\begin{eqnarray}
P_e &\le&  \sum_{\nu=r+1}^n\binom{n}{\nu}\varepsilon^{\nu}(1-\varepsilon)^{n-\nu}\nonumber\\
&+&  \sum_{\nu=1}^r
2^{\nu-r} \left( 1+ \frac{\binom{n-\nu}{K}}{\binom{n}{K}} \right)^{r}
 \binom{n}{\nu} \varepsilon^\nu(1-\varepsilon)^{n-\nu}  \;.
\label{bound_LDPC} 
\end{eqnarray}
\end{theorem}

\begin{IEEEproof} 
See Appendix~B.
\end{IEEEproof}

The same technique leads to the following bound for the Gallager ensemble of random LDPC codes.

\begin{theorem}\label{RankGal} The $(J,K)$-Gallager ensemble-average ML decoding error probability  for 
{$[n,n-r]$ codes,} $n=MK$, $r=MJ$,  and $M \gg 1$, is 
{upperbounded} by
\begin{eqnarray}
P_e &\le&  \sum_{\nu=1}^r  \sum_{\mu=0}^{J(n-\nu)/K}
 \min \left\{1,2^{\mu+\nu-r}\right\}
 \min\left\{1, 
\binom{\mu+J-1}{J-1}\binom{M}{\mu/J}^J \left( \frac{n-\nu}{n} \right)^{\mu K}  
 \right\}
   \binom{n}{\nu} \varepsilon^\nu(1-\varepsilon)^{n-\nu} \nonumber\\
   &+& \sum_{\nu=r+1}^n \binom{n}{\nu} \varepsilon^\nu(1-\varepsilon)^{n-\nu} \;.
\label{bound_Galn} 
\end{eqnarray}
\end{theorem}

\begin{IEEEproof} 
See Appendix~C.
 \end{IEEEproof}

We refer to the bounds (\ref{bound_LDPC}) and (\ref{bound_Galn})
as {\it R-bounds}, since they are based on estimating the rank of submatrices of $H$.   

Computations show that 
while for rates close to the capacity these bounds 
are rather tight,
for small $\varepsilon$ (or for rates 
{significantly lower} than the capacity)  these bounds are  weak for short codes. 
The reason for the bound untightness  is related to the Gallager ensemble properties 
discussed in detail in Appendix~A.

\section{Sliding-window near-ML decoding of QC LDPC codes} \label{sec_method}

In this section, we present  a new decoding algorithm for a class of QC LDPC codes. This class of codes is widely used in practice due to their extremely compact description. Moreover, under a so-called bi-diagonal structure restriction, QC LDPC codes have linear encoding complexity.
    
A binary QC LDPC block code  can be considered as a TB parent convolutional code determined by
a polynomial parity-check matrix whose entries are monomials or zeros.

A rate $R=b/c$ {\it parent}  LDPC convolutional code can be determined by
its polynomial parity-check matrix 
\begin{equation}
\arraycolsep=1.6pt \def\arraystretch{1.0}
H(D)=\left(\begin{array}{cccc}
h_{11}(D)&h_{12}(D)&\dots&h_{1c}(D)\\
h_{21}(D)& h_{22}(D)&\dots&h_{2c}(D)\\
\vdots&\vdots&\ddots&\vdots\\
h_{(c-b)1}(D)& h_{(c-b)2}(D)&\dots&h_{(c-b)c}(D)
\end{array}
\right) \; ,
\label{polyn}
\end{equation}
where $D$ is  a formal variable,
$h_{ij}(D)$ is  either zero or a monomial entry, that is, $h_{ij}(D)\in \{0,D^{w_{ij}}\}$
with $w_{ij}$ being a nonnegative integer, 
and $\mu=\max_{i,j} \{ w_{ij} \}$ is the syndrome memory.

The polynomial matrix (\ref{polyn}) determines an $[M_0c,M_0b]$  QC LDPC block code using a
set of polynomials  modulo $D^{M_0}-1$. If $M_0\to \infty$ we obtain an  LDPC convolutional code
which is considered as a parent convolutional code with respect to the QC LDPC block code for any finite $M_0$. By tailbiting the parent convolutional code to length $M_0 > \mu$, we obtain
the binary parity-check matrix
\begin{equation*}
\arraycolsep=3pt \def\arraystretch{1.2}
H_{\rm TB}=\begin{pmatrix}
H_{0}&H_{1}&\dots&H_{\mu-1}&H_{\mu}&{\bs 0}&\dots&\bs 0\\
{\bs 0}& H_{0}&H_{1}&\dots&H_{\mu-1}&H_{\mu}&\dots&\bs 0\\
\vdots &                 &\ddots               &\vdots&\vdots                     &\vdots                   &\ddots&\\
H_{\mu}& {\bs 0}&\dots&\bs 0&H_{0}&H_{1}&\dots&H_{\mu-1}\\
\vdots&\ddots&\vdots&\vdots&\vdots&\vdots&\vdots&\vdots\\
H_{1}&\dots&H_{\mu}&{\bs 0}&\dots&\bs 0&\dots&H_{0}
\end{pmatrix}
\label{tb}
\end{equation*}
of an equivalent (in the sense of column permutation) TB
code (all matrices $H_i$ including $H_{\rm TB}$ should have a transpose
operator to get the exact TB code \cite{JZ2015}),
where
$H_{i}$, $i=0,1,\dotsc,\mu$, are the binary $(c-b)\times c$ matrices in the series expansion
\begin{equation*}
	H(D)=H_{0}+H_{1}D+\cdots+H_{\mu}D^{\mu} \;.
\end{equation*}

If every column and row of $H(D)$ contains $J$ and $K$ nonzero entries, respectively,
we call $\mathcal{C}$ a $(J,K)$-regular  QC LDPC code and irregular otherwise.

Notice that by zero-tail termination \cite{JZ2015} of (\ref{polyn})  at length $W>\mu$, we can obtain a parity-check matrix of a
$[Wc,(W-\mu)b]$ ZT QC LDPC code.

Consider  a BEC with erasure probability $\varepsilon>0$.
Let $H$ be an {$M_0(c-b) \times M_0c$} parity-check matrix of a binary $[n=M_0c, k= M_0b,d_{\min}]$   QC LDPC block code, where $d_{\min}$ is the  minimum Hamming distance  of the code.  An ML decoder corrects any pattern of $\nu$ erasures if
$\nu\le  d_{\min}-1$.  If $d_{\min}\le \nu \le n-k $, then
a unique correct decision can be obtained for some erasure patterns. The number of such correctable patterns depends on the code structure.

%

As explained in Section~\ref{sec_prilim},  ML decoding over the BEC is equivalent to solving (\ref{eq_syndrome}). Furthermore, it is well-known that solving  a system of $t$ sparse linear equations  requires $O(t^2)$ operations \cite{wiedemann1986},\cite{lamacchia1990solving}. Typically, ML  decoding of  LDPC codes is performed in the form of hybrid decoding combining standard  iterative decoding with an additional decoding step applied to the result of the BP decoding in case it fails (see, for example, \cite{paolini2012maximum}, \cite{cunche2010analysis}).
The  computational complexity of hybrid decoding of LDPC codes over the BEC is of order  $\nu^2$, where $\nu \approx (1-R) n$. A thorough  complexity analysis of pivoting strategies for hybrid decoding of LDPC codes is presented in \cite{paolini2012maximum}.  It was shown in \cite{cunche2010analysis} that for a specified class of QC LDPC codes, the complexity of hybrid decoding can be reduced to $O(\nu\sqrt{\nu})$. This technique can be efficiently applied either to the decoding  of  short and moderate length QC LDPC codes or to the decoding of rather long high rate QC LDPC codes. In \cite{ kim2008efficient}, a near-linear time ML decoding algorithm  based on a simplified Gaussian elimination technique for  long (in the order of tens of thousands of  bits)  Raptor codes where  $r \ll n$ (code redundancy is about $1$ percent) was studied. However, if both $r$ and $n$ are   in the order of tens of thousands of bits, ML decoding of  LDPC  (or QC LDPC) codes over the BEC is  still considered computationally intractable. 

 In order to overcome this limitation, low-complexity  near-ML decoding algorithms were proposed in  \cite{pishro2004decoding}, \cite{vellambi2007}.  They are based on guessing codeword bits when standard BP decoding fails. This type of  decoding algorithms  requires a  number of trials of BP decoding equal to,  at least,   a polynomial function  of the number of  bit guesses. Notice that the number of bit guesses  grows with the channel erasure probability. The algorithm in \cite{vellambi2007} improves the bit guessing rule  compared to the one used in \cite{pishro2004decoding}. The suggested improvement is based on the observation that, typically, at lower channel erasure probabilities, the fraction of weight-two checks  among the unsatisfied checks is significant. This allows to simplify decoding by replacing the corresponding pairs of variables by one variable.  In \cite{pishro2004decoding} and  \cite{vellambi2007}, the suggested algorithms were evaluated and   compared  based only  on their simulated  BER performance.  
	
In this paper, we follow another approach, which takes into account the similarity of decoding techniques for convolutional codes and those of  QC block codes. The proposed algorithm resembles wrap-around suboptimal decoding of TB convolutional codes \cite{kudryashov1990decoding, shao2003two}. The main idea behind this approach is that Viterbi decoding is applied  to the wrapped-around trellis  of a TB code with all  initial state metrics equal to zero instead of running a separate Viterbi decoding from each initial state. This approach yields near-ML decoding performance with a linear number of decoding trials of Viterbi decoding. We apply this idea to QC LDPC codes.

 It is shown in \cite{lentmaier2010iterative} that iterative decoding thresholds of  regular LDPC  convolutional codes closely approach the ML decoding thresholds of the corresponding regular LDPC block codes.  However, this is not the case for block LDPC codes.  In order to improve the  decoding performance  for block LDPC codes,  we apply a sliding-window decoding algorithm 
which is modified for QC LDPC block  codes.  The suggested decoding algorithm is a hybrid decoding algorithm where  BP decoding of the QC LDPC code is followed by  so-called SWML decoding. The latter technique is applied “quasi-cyclically” to a relatively short sliding window, where the decoder performs ML decoding of a  ZT LDPC convolutional code. The decoding complexity of the proposed algorithm (simply referred to as SWML decoding in the sequel)  is polynomial (at most cubic)  in the window length, but only linear in the code length. Owing to the low computational complexity, the proposed hybrid decoding algorithm can be used for decoding codes of lengths in the order of hundreds of thousands of bits,  but its decoding  performance is limited by the maximum degree of the monomials  in the parity-check matrix of its parent LDPC convolutional code that in turn is restricted by the window size.

The SWML decoder is determined by a binary parity-check matrix 
 {\setlength{\arraycolsep}{4pt}
\begin{eqnarray}
H_{\rm W}=\left(\begin{array}{ccccccc}
H_{0}&\dots&H_{\mu}&{\bs 0}&\bs 0&\dots&{\bs 0} \\
{\bs 0}& H_{0}&\dots&H_{\mu}&\bs 0&\dots&{\bs 0}\\
\vdots&\ddots&\ddots&\   &\ddots&\ddots&\vdots\\
\bs 0&\dots &  \bs 0&  H_{0} &\dots&    H_{\mu}             &\bs 0\\
\bs 0&\bs 0&\dots& {\bs 0}&H_{0}&\dots&H_{\mu}\\
\end{array}
\right)
\label{zt}
\end{eqnarray}
of size $(W-\mu)(c-b)\times Wc$, where $W\ge 2\mu+1$ denotes the size of the decoding window in blocks.
The parity-check matrix (\ref{zt}) determines a ZT  LDPC parent convolutional  code.
We start decoding with   BP decoding applied to the original QC LDPC block 
code of length $n=M_0c$, and then apply ML decoding to  the ZT LDPC parent 
convolutional  code determined by the parity-check matrix (\ref{zt}).  
It implies solving a system of linear equations
\begin{equation*}
{ \bs x}_{I,\rm W} H^{\rm T}_{I,\rm W} =\bs s_{\rm W} \;,
\end{equation*}
where  ${\bs x}_{I,\rm W}=(x_{I,{\rm W},i},x_{I,{\rm W} ,i+1 \bmod n}, \dotsc, x_{I,{\rm W}, i+Wc-1 \bmod n})$,  $i=0,s, 2s, \dotsc  \bmod n$  is a subvector of ${\bs x}_{I}$ corresponding to the chosen window,  $s$ denotes the size of the window shift,  and ${\bs s}_{\rm W}$  and   $H_{I,\rm W}$ are the corresponding subvector of    ${\bs s}$  and  submatrix of $H_{I}$, respectively.  The final decision is made after $\alpha n/s$ steps, where $\alpha$ denotes the number of passes of sliding-window decoding.  The formal description of the decoding procedure is given below as Algorithms~1 and 2.


Notice that the choice of $s$ affects both the  performance and the complexity. By increasing $s$ we can speed up the decoding procedure at the cost of some performance loss. In the sequel, we use  $s=c$ bits that corresponds to the lowest possible FER.


\begin{algorithm}
\begin{algorithmic}[0]
\caption{ BP-BEC}
\While {there exist parity checks with only one erased  symbol} \\
{
Assign to the erased symbol the modulo-$2$ sum of all
nonerased symbols participating in the same parity check.
 }
\EndWhile
\end{algorithmic}
\end{algorithm}

\begin{algorithm}
\begin{algorithmic}[0]
\caption{\label{WAD}Wrap-around  algorithm for near-ML
decoding of QC LDPC codes over the BEC}
\Statex {\bf Input:} BEC output $\bs y$.
\Statex Perform BP decoding of $\bs y$.
\Statex ${\rm wstart} \gets 0$; 
\Statex ${\rm wend} \gets W-1$; 
\Statex $\rm corrected \gets 1$;
\While {$\rm corrected>0$}
\Statex {\hspace{5mm}$\rm corrected \gets 0$;}%
\Statex {\hspace{4mm} Apply ML decoding to the window $(y_{\rm wstart},\dotsc,y_{\rm { wend}})$;}%
\Statex {\hspace{4mm} ${\rm wstart \gets {\rm wstart}}+s \bmod n$;}%
\Statex {\hspace{4mm} ${\rm wend \gets \rm wend}+s  \bmod n$;}%
\If {$\rm wstart = 0$} 
{
 \Statex \hspace{10mm}{corrected $\gets $  number of corrected erasures after a full round;}%
}
	\EndIf
\EndWhile
\Statex {{\bf return} $\bs y$}
\end{algorithmic}
\end{algorithm}
Next, we present an example of selecting window positions in the parity-check matrix $H$ of the QC LDPC code.  The window size in blocks is equal to $(W-\mu) \times W =(7-3)\times 7$. The two first and three last window positions in each pass are shown.  

\begin{example}

\[\small
H = \left( \: 
\begin{array}{*{8}{c}}
\cline{1-7}
\multicolumn{1}{|c}{H_0} & H_1 & H_2   & H_3 &\bf{ 0} &\bf{0} & \multicolumn{1}{c|}{\bf{0}}&\bf{0}\\
\multicolumn{1}{|c}{\bf{0}} &H_ 0 &H_1    &H_2 &H_3 & \bf{0} &\multicolumn{1}{c|}{\bf{0}}&\bf{0}\\
\multicolumn{1}{|c}{\bf{0}} & \bf{0} &H_0  &H_ 1 & H_2 &H_3 &  \multicolumn{1}{c|}{\bf{0}}&\bf{0}\\ 
\multicolumn{1}{|c}{\bf{0}} & \bf{0} &\bf{0}  &H_ 0 & H_1 &H_2 &  \multicolumn{1}{c|}{H_3}&\bf{0}\\
\cline{1-7}
{\bf 0}&{\bf 0}&{\bf 0}&{\bf 0}&H_{0}&H_{1}&H_{2}&H_{3}\\
H_{3}&{\bf 0}&{\bf 0}&{\bf 0}&{\bf 0}&H_{0}&H_{1}&H_{2}\\
H_{2}&H_{3}&{\bf 0}&{\bf 0}&{\bf 0}&{\bf 0}&H_{0}&H_{1}\\
H_{1}&H_{2}&H_{3}&{\bf 0}&{\bf 0}&{\bf 0}&{\bf 0}&H_{0}\\
\end{array}
\right)\; , \] 

\[\small
H = \left( \: 
\begin{array}{*{8}{c}}
H_0 & H_1 & H_2   & H_3 &\bf{ 0} &\bf{0} &\bf{0}&\bf{0}\\
\cline{2-8}
\bf{0} &\multicolumn{1}{|c}{H_ 0} &H_1    &H_2 &H_3 & \bf{0} &\bf{0}&\multicolumn{1}{c|}{\bf{0}}\\
\bf{0} &\multicolumn{1}{|c}{ \bf{0}} &H_0  &H_ 1 & H_2 &H_3 & \bf{0}&\multicolumn{1}{c|}{\bf{0}}\\ 
\bf{0} & \multicolumn{1}{|c}{\bf{0}} &\bf{0}  &H_ 0 & H_1 &H_2 & H_3&\multicolumn{1}{c|}{\bf{0}}\\
{\bf 0}&\multicolumn{1}{|c}{{\bf 0}}&{\bf 0}&{\bf 0}&H_{0}&H_{1}&H_{2}&\multicolumn{1}{c|}{H_{3}}\\
\cline{2-8}
H_{3}&{\bf 0}&{\bf 0}&{\bf 0}&{\bf 0}&H_{0}&H_{1}&H_{2}\\
H_{2}&H_{3}&{\bf 0}&{\bf 0}&{\bf 0}&{\bf 0}&H_{0}&H_{1}\\
H_{1}&H_{2}&H_{3}&{\bf 0}&{\bf 0}&{\bf 0}&{\bf 0}&H_{0}\\
\end{array}
\right)\; , \] 

\[\vdots\]

\[\small
H = \left( \: 
\begin{array}{*{8}{c}}
\multicolumn{1}{|c}{H_0} & H_1 & H_2   &\multicolumn{1}{c|}{ H_3} &\bf{ 0} &\multicolumn{1}{|c}{\bf{0}} &\bf{0}&\multicolumn{1}{c|}{\bf{0}}\\
\cline{1-4}
\cline{6-8}
\bf{0} &H_ 0 &H_1    &H_2 &H_3 & \bf{0} &\bf{0}&\bf{0}\\
\bf{0} & \bf{0} &H_0  &H_ 1 & H_2 &H_3 & \bf{0}&\bf{0}\\ 
\bf{0} &\bf{0} &\bf{0}  &H_ 0 & H_1 &H_2 & H_3&\bf{0}\\
\bf{ 0}&\bf{ 0}&\bf{ 0}&\bf{ 0}&H_{0}&H_{1}&H_{2}&H_{3}\\
\cline{1-4}
\cline{6-8}
\multicolumn{1}{|c}{H_{3}}&{\bf 0}&{\bf 0}&\multicolumn{1}{c|}{\bf{ 0}}&\bf{0}&\multicolumn{1}{|c}{H_{0}}&H_{1}&\multicolumn{1}{c|}{H_{2}}\\
\multicolumn{1}{|c}{H_{2}}&H_{3}&{\bf 0}&\multicolumn{1}{c|}{\bf{0}}&{\bf{ 0}}&\multicolumn{1}{|c}{\bf{ 0}}&H_{0}&\multicolumn{1}{c|}{H_{1}}\\
\multicolumn{1}{|c}{H_{1}}&H_{2}&H_{3}&\multicolumn{1}{c|}{\bf {0}}&\multicolumn{1}{c|}{\bf{ 0}}&{\bf{0}}&{\bf{ 0}}&\multicolumn{1}{c|}{H_{0}}\\
\end{array}
\right)\; , \]

\[\small
H = \left( \: 
\begin{array}{*{8}{c}}
\multicolumn{1}{|c}{H_0} & H_1 & H_2   &{ H_3} &\multicolumn{1}{c|}{\bf{ 0}} &{\bf{0}} &\multicolumn{1}{|c}{\bf{0}}&\multicolumn{1}{c|}{\bf{0}}\\
\multicolumn{1}{|c}{\bf{0}} &H_ 0 &H_1    &H_2 &\multicolumn{1}{c|}{H_3 }& \bf{0} &\multicolumn{1}{|c}{\bf{0}}&\multicolumn{1}{c|}{\bf{0}}\\
\cline{1-5}
\cline{7-8}
\bf{0} & \bf{0} &H_0  &H_ 1 & H_2 &H_3 & \bf{0}&\bf{0}\\ 
\bf{0} &\bf{0} &\bf{0}  &H_ 0 & H_1 &H_2 & H_3&\bf{0}\\
\bf{ 0}&\bf{ 0}&\bf{ 0}&\bf{ 0}&H_{0}&H_{1}&H_{2}&H_{3}\\
H_{3}&{\bf 0}&{\bf 0}&\bf{ 0}&\bf{0}&H_{0}&H_{1}&H_{2}\\
\cline{1-5}
\cline{7-8}
\multicolumn{1}{|c}{H_{2}}&H_{3}&{\bf 0}&{\bf{0}}&{\bf{ 0}}&\multicolumn{1}{|c}{\bf{ 0}}&\multicolumn{1}{|c}{H_{0}}&\multicolumn{1}{c|}{H_{1}}\\
\multicolumn{1}{|c}{H_{1}}&H_{2}&H_{3}&{\bf {0}}&\multicolumn{1}{c|}{\bf{ 0}}&{\bf{0}}&\multicolumn{1}{|c}{\bf{ 0}}&\multicolumn{1}{c|}{H_{0}}\\
\end{array}
\right)\; , \]

\[\small
H = \left( \: 
\begin{array}{*{8}{c}}
\multicolumn{1}{|c}{H_0} & H_1 & H_2   &{ H_3} &{\bf{ 0}} &\multicolumn{1}{c|}{\bf{0}} &\multicolumn{1}{c|}{\bf{0}}&\multicolumn{1}{c|}{\bf{0}}\\
\multicolumn{1}{|c}{\bf{0}} &H_ 0 &H_1    &H_2 &{H_3 }&\multicolumn{1}{c|}{ \bf{0}} &\multicolumn{1}{c|}{\bf{0}}&\multicolumn{1}{c|}{\bf{0}}\\
\multicolumn{1}{|c}{\bf{0}} & \bf{0} &H_0  &H_ 1 & H_2 &\multicolumn{1}{c|}{H_3} & \multicolumn{1}{c|}{\bf{0}}&\multicolumn{1}{c|}{\bf{0}}\\ 
\cline{1-6}
\cline{8-8}
\bf{0} &\bf{0} &\bf{0}  &H_ 0 & H_1 &H_2 & H_3&\bf{0}\\
\bf{ 0}&\bf{ 0}&\bf{ 0}&\bf{ 0}&H_{0}&H_{1}&H_{2}&H_{3}\\
H_{3}&{\bf 0}&{\bf 0}&\bf{ 0}&\bf{0}&H_{0}&H_{1}&H_{2}\\
{H_{2}}&H_{3}&{\bf 0}&{\bf{0}}&{\bf{ 0}}&{\bf{ 0}}&{H_{0}}&{H_{1}}\\
\cline{1-6}
\cline{8-8}
\multicolumn{1}{|c}{H_{1}}&H_{2}&H_{3}&{\bf {0}}&{\bf{ 0}}&\multicolumn{1}{c|}{\bf{0}}&\multicolumn{1}{c|}{\bf{ 0}}&\multicolumn{1}{c|}{H_{0}}\\
\end{array}
\right)\;. \]	
	
\end{example}

\section{Numerical results and simulations \label{numres}}

\subsection{Short codes}

In 
Fig. \ref{BEC_bounds4}, we compare  upper and lower bounds 
on the error probability of ML decoding of  
short codes on the BEC.  First, notice that the  lower bounds in Theorems~\ref{triv} and \ref{lower} (sphere packing and the tightened 
sphere packing bound)  almost coincide with each other near the channel capacity. However, 
 the new bound is significantly tighter than the known one in the low erasure probability region. 
For further comparisons we use the new bound whenever information on the code minimum distance is
available.   The upper bounds in Theorems~\ref{RankLdpc} and \ref{RankGal} are also 
presented in  Fig. \ref{BEC_bounds4}. 
These two bounds are indistinguishable at high symbol erasure probabilities
but the difference between them  is visible in the low $\varepsilon$ region where all bounds are rather weak.  Notice that for high $\varepsilon$ random coding  bounds for LDPC codes are close to
those of random linear codes. For $(J,K)$-regular  LDPC codes with  $J=4$, both S and R-bounds are  almost as good as the random bound for general linear codes  of the same length and dimension in a wide range of  $\varepsilon$ values.   
{
In Fig. \ref{BEC_bounds4},  as well as in subsequent figures, S-bounds
(\ref{gen_spectr})
are computed based on  the ensemble average spectra of the corresponding $(J,K)$-regular Gallager code ensembles {by using the recurrent procedure described  in Section \ref{sec_spec}}.


 \begin{figure}
\begin{center}
\includegraphics[width=90mm]{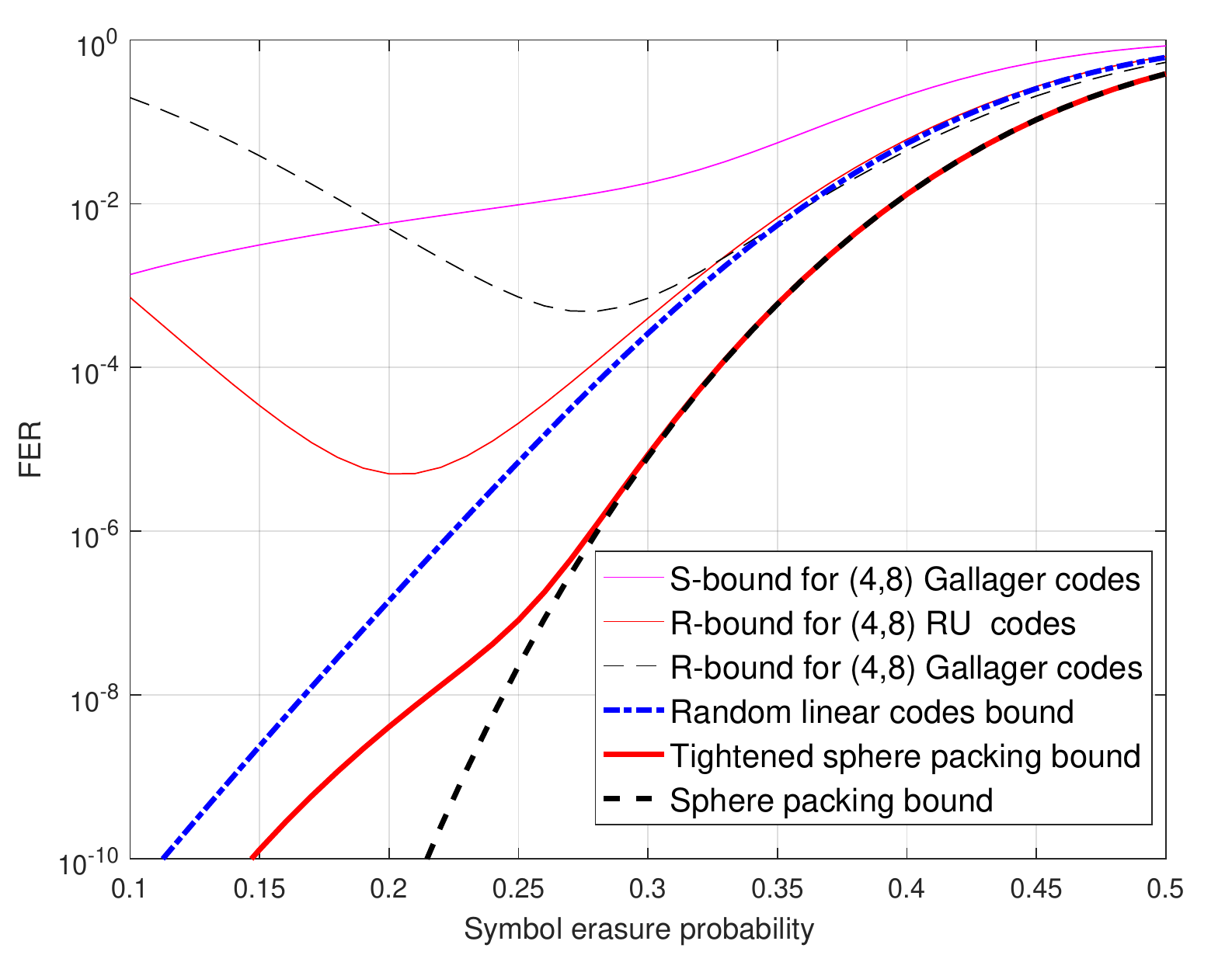}   
\caption{\label{BEC_bounds4}  Bounds for binary $(4,8)$-regular LDPC codes of length $n=96$ and rate $R=1/2$.
The 
S-bound is defined by (\ref{gen_spectr}), and R-bounds for RU and Gallager codes 
are defined by (\ref{bound_LDPC}) 
and (\ref{bound_Galn}), respectively.
The random coding bound is computed by (\ref{genbound})--(\ref{precise}), while the  sphere packing bound and 
the tightened sphere packing bound are computed according to (\ref{lower_b}) and (\ref{LT}), respectively.   
  }
\end{center}
\end{figure}

In Fig.~\ref{BEC_short}, we compare  upper bounds on the ML decoding FER performance  for 
the  Gallager  ensemble of $(J,K)$-regular
LDPC codes  with different pairs $(J,K)$ to the upper bounds for general linear codes of different code rates.  Interestingly, the convergence rate 
of the bounds for LDPC codes to the bounds for linear codes depends on the code rate. For rate $R=1/3$,    
even for rather sparse codes with column weight $J=4$, their  FER performance under ML decoding
is very close to the FER performance of general linear codes.

\begin{figure}
\begin{center}
\includegraphics[width=130mm]{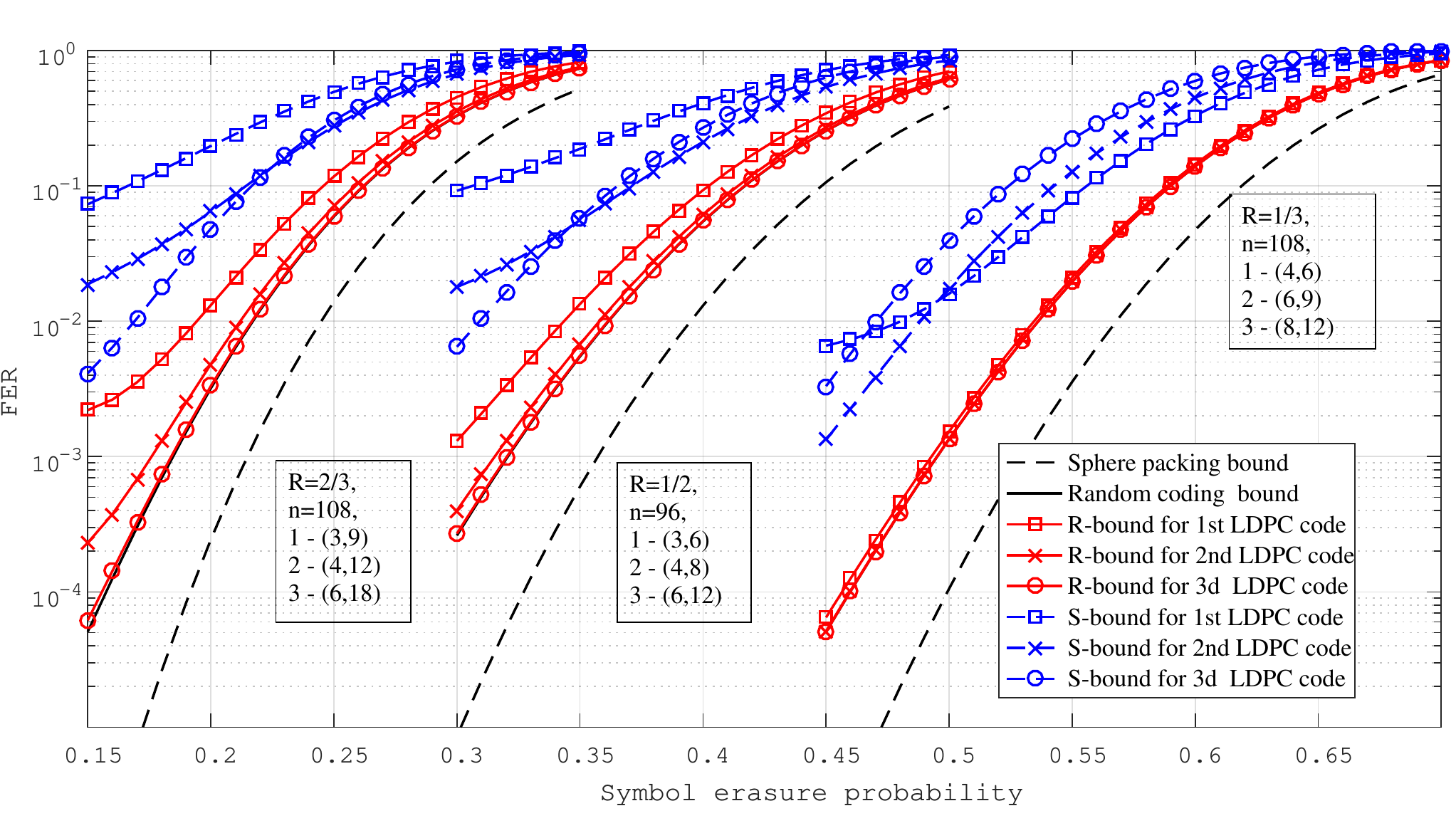}   
\caption{\label{BEC_short}  Error probability bounds for  the binary $(J,K)$-regular Gallager  LDPC codes of length $n\approx 100$.
The 
S and R-bounds are defined by (\ref{gen_spectr}) and  (\ref{bound_Galn}), respectively.
Random coding bound for linear codes   is computed by (\ref{genbound})--(\ref{precise}). The lower bound is (\ref{LT}).   
  }
\end{center}
\end{figure}

 Next, we present simulation results for two ensembles of LDPC codes. 
In Fig.~\ref{Comp36vs48}, we compare  the best code among $10$ randomly selected  $(3,6)$-regular LDPC codes and the best code among $10$ randomly selected  $(4,8)$-regular LDPC codes of the two ensembles. 
As predicted by bounds, the ML decoding performance of the  $(4,8)$-regular codes is much better than that of the 
$(3,6)$-regular  codes in both ensembles. Notice that  codes 
from the Gallager ensemble are weaker than the RU codes with the same parameters. This can be explained by  the rate bias, 
the code rate for the  RU codes is typically equal to $R=1/2$, whereas for the Gallager codes  the rate is $R\ge 52/96=0.5417$.
A  second  reason to the difference in the FER performance   is the  approximation discussed in Appendix~A.
The performance of the RU codes perfectly matches the R-bound.  Moreover, the best $(4,8)$-RU code shows even better FER performance than  the average FER performance of general linear codes in the  high erasure probability region.

\begin{figure}
\begin{center}
\includegraphics[width=90mm]{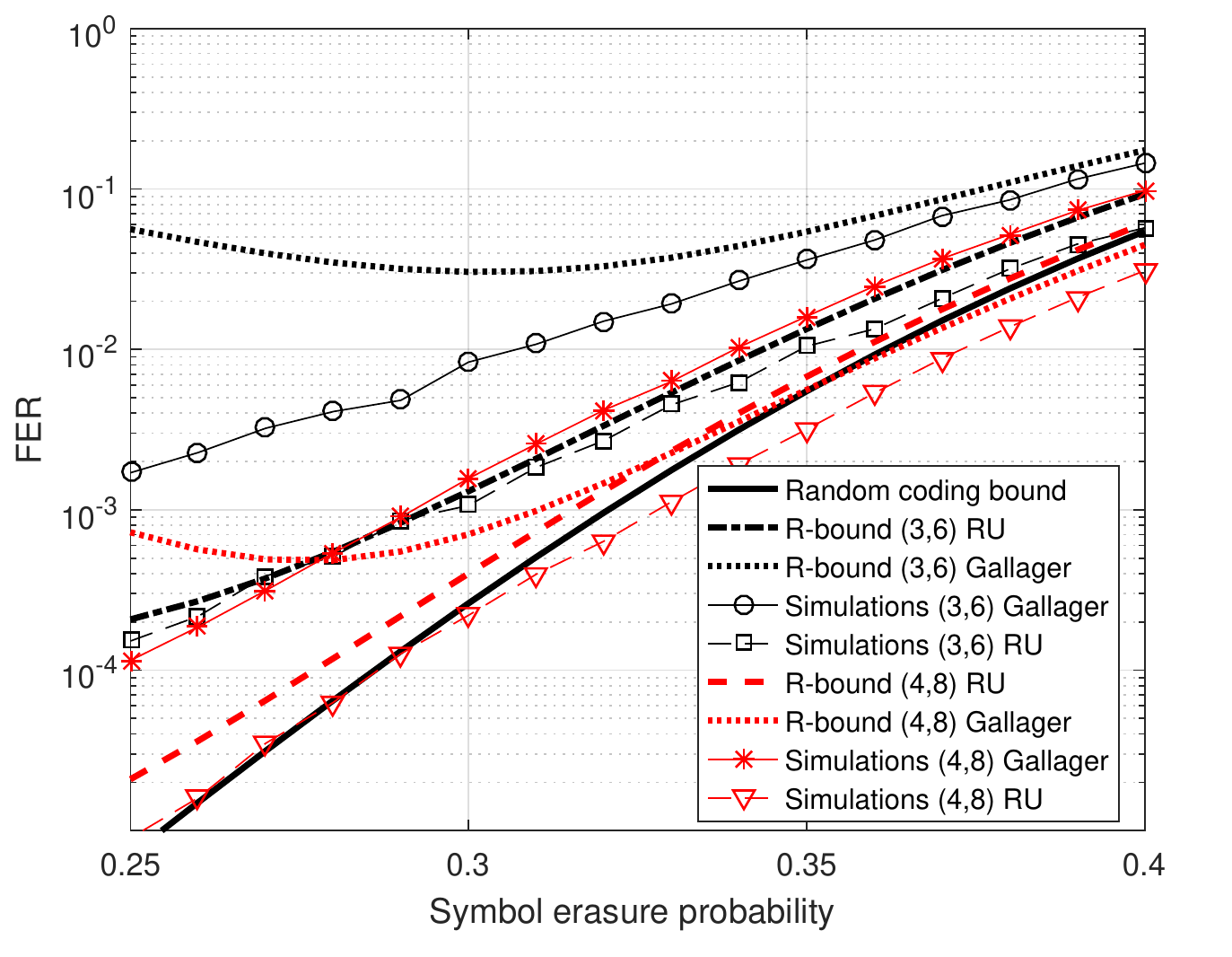}   
\caption{\label{Comp36vs48}  
Simulation results for $(3,6)$-regular and $(4,8)$-regular codes of length   $n=96$ from the  Gallager and RU ensembles.
The S-bound is defined  by (\ref{gen_spectr}),   and R-bounds  for RU and Gallager codes  are defined 
by (\ref{bound_LDPC}) and (\ref{bound_Galn}), respectively. 
The random coding bound for linear codes  is computed by (\ref{genbound})--(\ref{precise}).
  }
\end{center}
\end{figure}


\subsection{Codes of moderate length}

The FER performance for relatively long codes of length $n=1008$
is shown in Fig.~\ref{BEC_long}. 
Notice that the difference between the lower and upper bounds for general linear codes, which was 
noticeable for short codes becomes very small for $n=1008$. 
Since the lower bound   (\ref{lower_b}) is simply the probability of more than $r$ erasures,
the fact that the upper and the lower bounds almost coincide  leads us to the conclusion that even for rather low channel erasure probabilities
$\varepsilon$, achieving ML decoding performance requires  correcting 
almost all combinations of  erasures of weight close to the code redundancy $r$.  Notice that according to the S-bounds in Fig.~\ref{BEC_long}, error floors are expected in the low erasure 
probability region.  The error-floor level strongly depends on $J$ and $K$, and rapidly decreases with increasing $J$.

\begin{figure}
\begin{center}
\includegraphics[width=130mm]{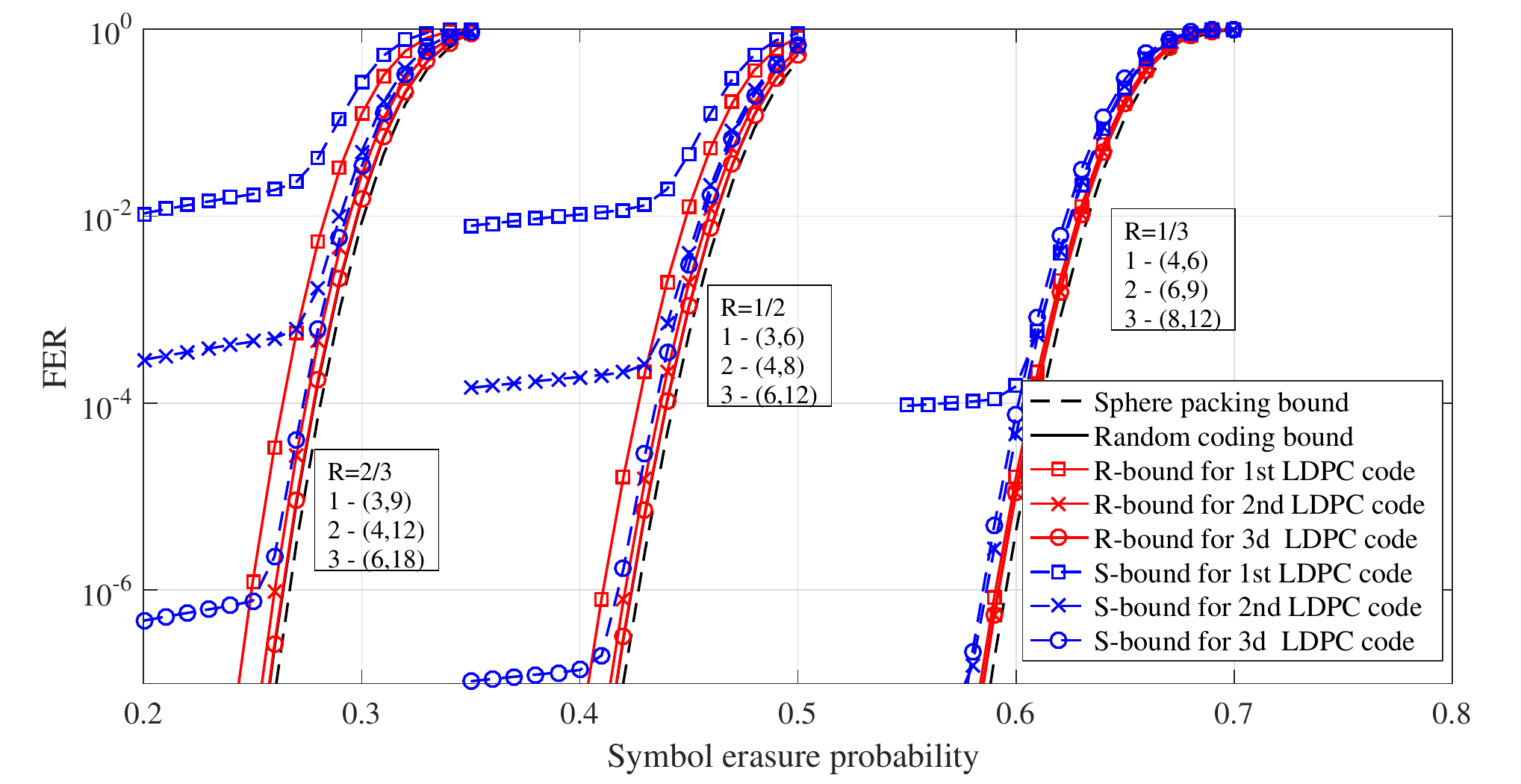}   
\caption{\label{BEC_long}  Error probability bounds for the binary $(J,K)$-regular Gallager LDPC codes of length $n= 1008$.
S and  R-bounds are defined by (\ref{gen_spectr}) 
and  (\ref{bound_Galn}), respectively. 
The random coding bound for  linear codes  is computed by (\ref{genbound})--(\ref{precise}).  The lower bound is  (\ref{LT}).   
  }
\end{center}
\end{figure}


\begin{figure}
\begin{center}
\includegraphics[width=90mm]{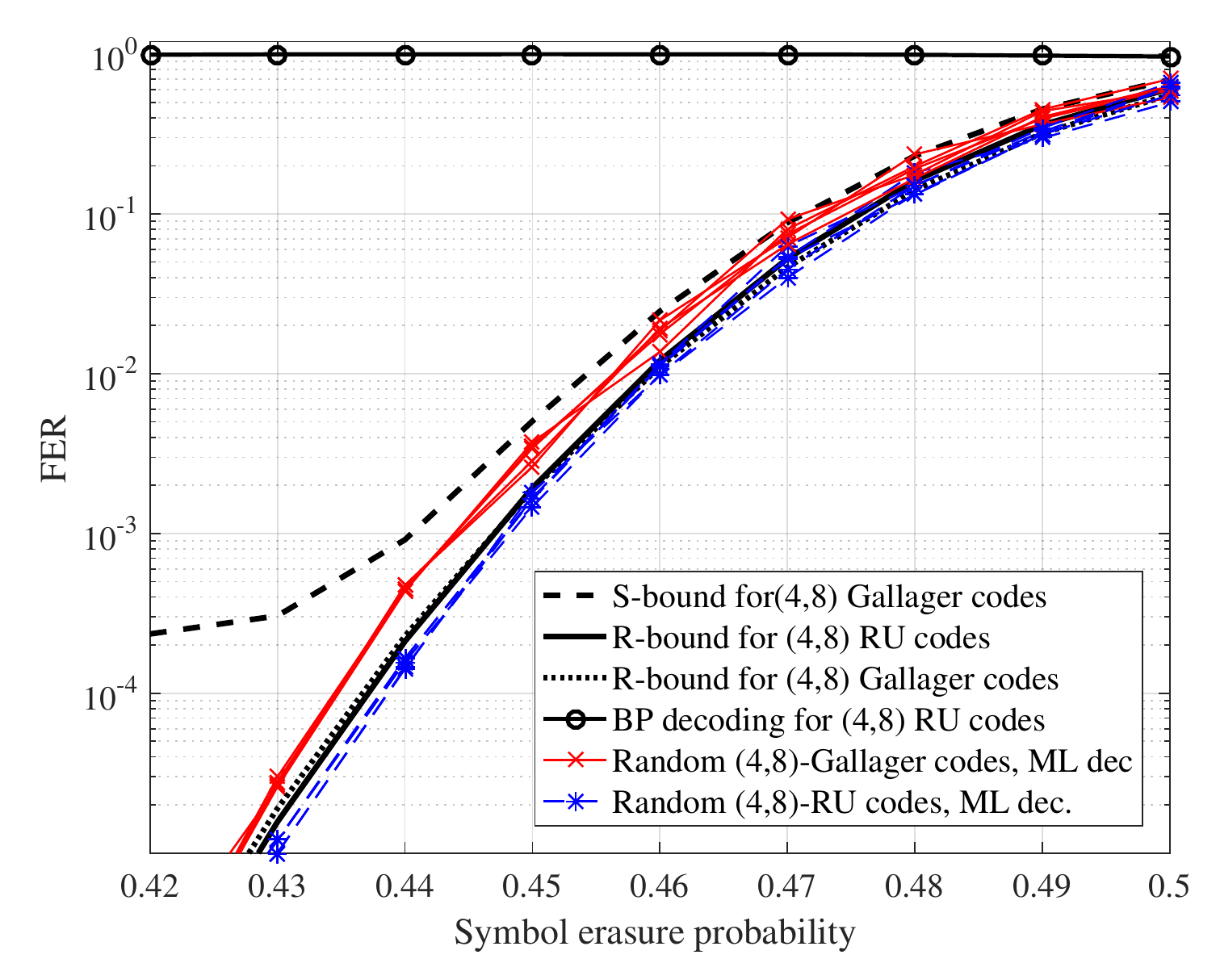}   
\caption{\label{BEC48_1000}  Error probability bounds and simulation results for 
$(4,8)$-regular LDPC codes of 
 length $n=1008$.
 The 
S-bound is defined by (\ref{gen_spectr}), and R-bounds for the RU and Gallager codes 
are defined by (\ref{bound_LDPC}) 
and (\ref{bound_Galn}), respectively.
   }
\end{center}
\end{figure}

In Fig. \ref{BEC48_1000}, we show simulation results for $5$ randomly selected 
 $(4,8)$-regular codes of length $n=1008$, respectively, from the Gallager and RU ensembles. Notice that the simulated 
performance for $5$ randomly selected $(3,6)$-regular codes of the same length demonstrates  similar behaviour and for this reason is not presented in the plot.  In the same 
plots the rank and spectral bounds for the corresponding  ensembles are shown. 
We 
{observe} that for rates close to the capacity, the rank and spectral  bounds demonstrate approximately the same behavior.  For low channel erasure probabilities the spectral bound predicts  error floors.
As expected, the spectral bound is weak for  rates far below the channel capacity. 
%
Since all $10$ 
codes ($5$ for the Gallager ensemble and $5$ for the RU ensemble) show identical FER performance, we present only one of the BP decoding FER curves in the figure. 

\subsection{Sliding-window  near-ML  decoding of  QC LDPC codes}

We compare  the FER performance of BP and SWML decoding  for three QC LDPC  codes of length $M_{0}c=4800$ to the S and R upper bounds on the ML decoding performance. The parameters  of the codes, including the minimum distance $d_{\min}$ and the stopping distance $d_{\rm stop}$ of the QC LDPC code of length $Wc$, where $W$ is  the window size in blocks,  and  the SWML decoder parameters are summarized in Table~\ref{Table_WML}. The irregular rate $R=12/24$  QC LDPC code  is determined by the monomial  parity-check matrix of its parent convolutional code optimized by the technique described in \cite{Boch2016}. This matrix  has the form
\begin{equation}H(D)=\left(
\begin{array}{cc}
H_{\rm{bd}}(D) & H_{\rm{a}}(D)
\end{array}
\right)\; ,
\label{parent}
\end{equation}
where $H_{\rm{bd}}$ is a bidiagonal matrix of size $12 \times 11$  with ones on the diagonals and zeros elsewhere, and $H_{\rm a}$ is a $12 \times 13$ matrix whose {\em  degree matrix} is
\begin{equation} \label{eq:Ha}
\small
\arraycolsep=2.5pt \def\arraystretch{0.9}
  \begin{pmatrix}
\begin{array}{rrrrrrrrrrrrr}
       0&  -1 & -1&  -1&   0&  -1 &  0&  -1&  -1&  -1&  -1&   0 &  0\\
      21&  -1 & -1&  -1 & -1 &  0&  -1&  -1 & -1 &  0&  -1&  -1&   3 \\
    -1 & -1  &-1  & 0 &  2  &-1 & -1&  -1&  -1&  -1&   0&   5 & -1 \\
    -1 & -1 &  0  &-1 & -1&  -1&  -1&  -1 &  0&  10&  -1&   7 & -1\\ 
    -1&  -1 & -1&  14 & -1&  -1 & -1 &  0&  -1&  -1 & 10&  -1 & 15 \\
    -1 &  0 & -1 & -1 & -1 & 20 & -1 & -1 &  7&  -1 & -1 & 11&   3 \\
    -1 & -1 &  6 & -1 &  4 & -1&  -1 & -1 & -1 &  9 & -1 & -1 &  0 \\
    -1&  18&  -1&  -1 & -1&  -1&  13&  -1&  -1 & -1 &  7&  19&  14\\ 
    -1 & -1&  -1 & 21&  -1 &  3 & -1 & -1&  -1 & 11&  -1 & 17&  13 \\
    -1 & 21&  -1 & -1 & -1 & -1 & -1 & 19&  -1&   2&  -1 & 12 & 14 \\
     0 & -1 & -1 & -1  &-1 & -1 & -1 & -1 & 18 & -1&  11&   2&   1 \\
     -1 & -1&  12 & -1&  -1&  -1 &  5 & 15&  -1 & -1 & 17 &  9&  -1
\end{array}
\end{pmatrix} \, .
\end{equation}
The polynomial parity-check matrix $H(D)$ is obtained from this degree matrix by replacing each negative entry with a zero, and replacing each nonnegative entry $e$ with $D^e$. 

A  regular rate $R=3/6$ QC LDPC code   determined by the parity-check matrix of \cite[Table~IV]{bocharova2012searching}
and a regular rate $R=8/16$ double-Hamming (DH) based QC LDPC code \cite{bocharova2011double}   determined by the degree matrix}
\[
\small
\arraycolsep=2.5pt \def\arraystretch{0.9}
  \begin{pmatrix}
\begin{array}{rrrrrrrrrrrrrrrrrr}
   11 &-1&   4 & -1& -1&  -1&  -1&  5&    6&   5&  -1& -1& 15& -1& 3& 11\\
    13& -1&  -1&  11& -1&  -1&  1&   -1&   13&  2&  -1&  2& -1& 13& -1& 7\\
    3&  -1&  -1&  -1&  5&  -1&  10&  9&    -1&  3&   8& -1& -1& 10& 8& -1\\
    12& -1&  -1&  -1& -1&  1&   8&   15&   9&  -1&   8&  4&  3& -1& -1& -1\\
    -1&  13& -1&   1& 10&  12&  2&   -1&   -1& -1&  -1&  0& -1& 6&  6& -1\\
    -1&  11&  6&  -1&  2&  6&   -1&   14&  -1& -1&  14& -1&  0& -1& 2& -1\\
    -1 & 13&  3&  5&  -1&  5&   -1&   -1&  10& -1&  -1& -1& 10& 12& -1& 7\\
    -1  &3  & 4 & 10&  4&  -1&  -1&   -1&  -1&  3&  14& 11& -1& -1& -1& 4
\end{array}
\end{pmatrix} 
\]
were simulated as well.  The minimum and stopping distances of the codes (of length $Wc$) are computed by using the algorithm in \cite{rosnes2009efficient}, \cite{rosnes2012addendum}. 
Simulation results and bounds  are presented in Fig.~\ref{WML}.

The SWML decoding FER performance of the DH code is the best among the SWML decoding FER performances of the simulated codes and it is   close to the theoretical bounds, despite its  very low decoding complexity.  In contrast, the FER performance of BP decoding of this code is extremely poor. This is not surprising, since  on one hand, the DH  code has better distance properties among the three codes, and on the other hand, its parity-check matrix is rather dense. This improves its ML decoding performance compared to the other codes and also makes it less suitable for iterative decoding.   For the $(3,6)$-regular QC LDPC code, as well as for the irregular QC LDPC code, BP decoding performs much better than for the DH code, but the   FER performance 
of SWML decoding  of these codes is worse  than that of the DH code. Notice that the considered irregular code has average column weight of $3.125$, which makes the comparison with the R and S-bounds for $(3,6)$ and $(4,8)$-regular codes fair. The gap between the FER performance of BP and SWML decoding of these two codes is not large.  The SWML decoding performance of the $(3,6)$ regular code is worse than that of BP decoding of the irregular code despite that the minimum distance of the $(3,6)$-regular code is larger. 
The SWML decoding performance of the $(3,6)$-regular code mimics the behavior of the S-bound. 

In Fig. \ref {SWML_ML},  we compare the FER performance of BP and ML decoding  for the  irregular QC LDPC code of length $4800$ with the FER performance of SWML decoding  with different window sizes for the same code. It is easy to see that  the FER performance of SWML decoding is approaching the FER performance of ML decoding  with increasing window size. The ML decoding curve is obtained by running an inactivation decoder \cite{Shokrollahi2005}.

In Figs. \ref{c_th} and \ref{c_th_fer}, we compare the simulated FER and BER performance of SWML decoding of three QC LDPC codes of length of a few tens of thousands of bits $(M_{0}=4000-10000)$  with the BP and ML decoding thresholds computed  by the density evolution analysis
\cite{richardson2008modern}.  Codes of rate $3/6$ and $8/16$ were obtained by increasing $M_{0}$ for the code in \cite[Table~IV]{bocharova2012searching} and the DH code, respectively. The irregular rate-$12/24$ code is  determined by the parity-check matrix  (\ref{parent}) with  $ H_{\rm{a}}(D)$ given by 
\[
\small
\arraycolsep=2.5pt \def\arraystretch{0.9}
  \begin{pmatrix}
\begin{array}{rrrrrrrrrrrrr}
  0& -1& -1&  0& -1& -1&  0& -1& -1& -1&  0&  0&  0\\
 -1& -1 &-1& -1& -1& -1& -1&  0& -1&  0 &-1&  5& 21\\
 -1&  0 &-1 &-1 & 0& -1 &-1& -1&  0& -1 &-1& 29& 41\\
 -1& -1&  0 &-1 &-1 &-1& -1& 81& -1& -1&  0& 50& 54\\
  55& 72& -1& -1& -1&  0& -1& -1& -1& -1& -1& 85&  5\\
 -1 &-1 &-1& 46 &-1& -1& -1& -1& 71& 29& -1 &23& 24\\
 -1 &-1 & 89 &-1 &-1& 50& -1& -1& -1& -1& -1 &93& 78\\
 -1 &-1 &-1 &-1 &36& -1& -1& -1 &-1 &50& 32&  8 &90\\
  -1& -1 &-1 &-1 &-1& -1& 56& -1 &26& -1 &-1&  9& 51\\
  -1& 94& -1 &69 &-1 &89& -1& -1 &-1 &-1 &-1 &84& 98\\
   0 &-1 &-1& -1 &-1& -1 &-1 &29& -1& -1& -1 &85&  8\\
  -1 &-1& 41& -1& 92& -1& 13& -1& -1 &-1 &-1& 39& 83
\end{array}
\end{pmatrix}. 
\]
Parameters of the codes together with the parameters of the SWML decoder are presented in Table \ref{Table_WML_long}. 
It follows from the presented curves that the SWML decoder significantly outperforms the BP decoder in the near-capacity region. 
Interestingly enough,  unlike for regular codes, the BP decoding threshold for irregular codes computed by density evolution is much lower than the simulated achievable performance.  However, due to memory restrictions,
the SWML decoder does not achieve ML decoding performance 
of ensemble-average regular  LDPC codes of the same length.  Moreover,  the decoding thresholds are obtained by the density evolution approach which, as well as the bounds presented in Section \ref{sec_bound}, do not take into account restrictions imposed by the quasi-cyclic structure of the codes. This fact increases the gap  between the thresholds and the simulated performance.

In Fig. \ref{comp_guess}, we compare the simulated FER and BER performance of the irregular $R=12/24$ QC LDPC  code of length $4800$ determined by the parity-check matrix (\ref{parent}) with $H_{\rm a}(D)$ given by (\ref{eq:Ha}) and lifting degree $M_{0}=200$ under SWML decoding with a window size of $W=51$ blocks and the BER performance of random LDPC codes of length $10^{4}$ decoded by the two bit guessing based  decoding algorithms ($A$ and $C$) in \cite{pishro2004decoding}. Although the window
size of the SWML decoder is much smaller than the code length and the code is two times shorter than the irregular codes in  \cite{pishro2004decoding}, the difference in performance is noticeable. 

\begin{figure}
\begin{center}
\includegraphics[width=100mm]{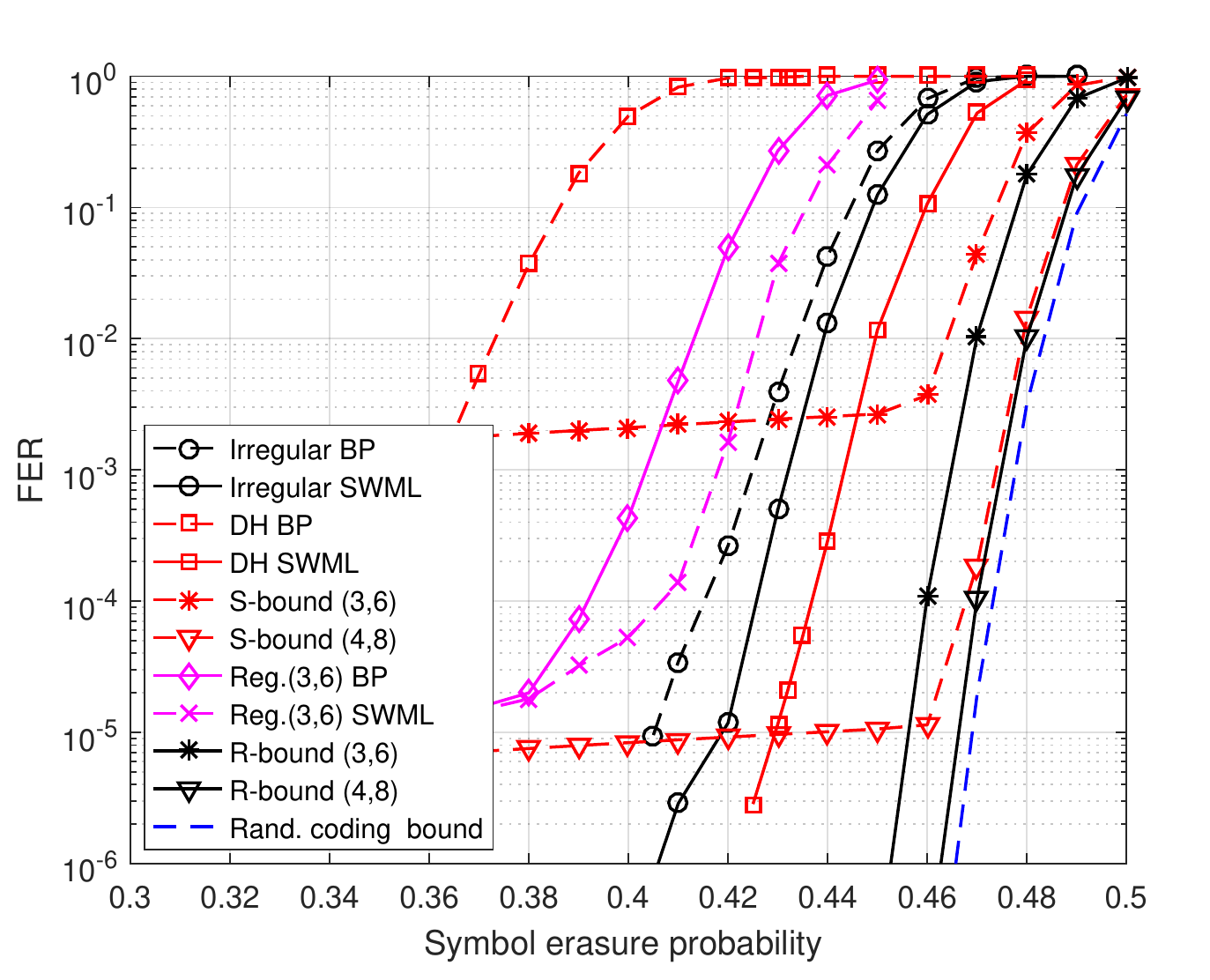}   
\caption{\label{WML}  Error probability bounds and the simulated FER performance of  
SWML decoding of QC LDPC codes of length  $n=4800$.
S and  R-bounds are defined by (\ref{gen_spectr}) 
and (\ref{bound_LDPC}), respectively. 
The average FER performance of ML decoding  for random  linear codes is computed by (\ref{genbound})--(\ref{precise}).
  }
\end{center}
\end{figure}

\begin{figure}
\begin{center}
\includegraphics[width=100mm]{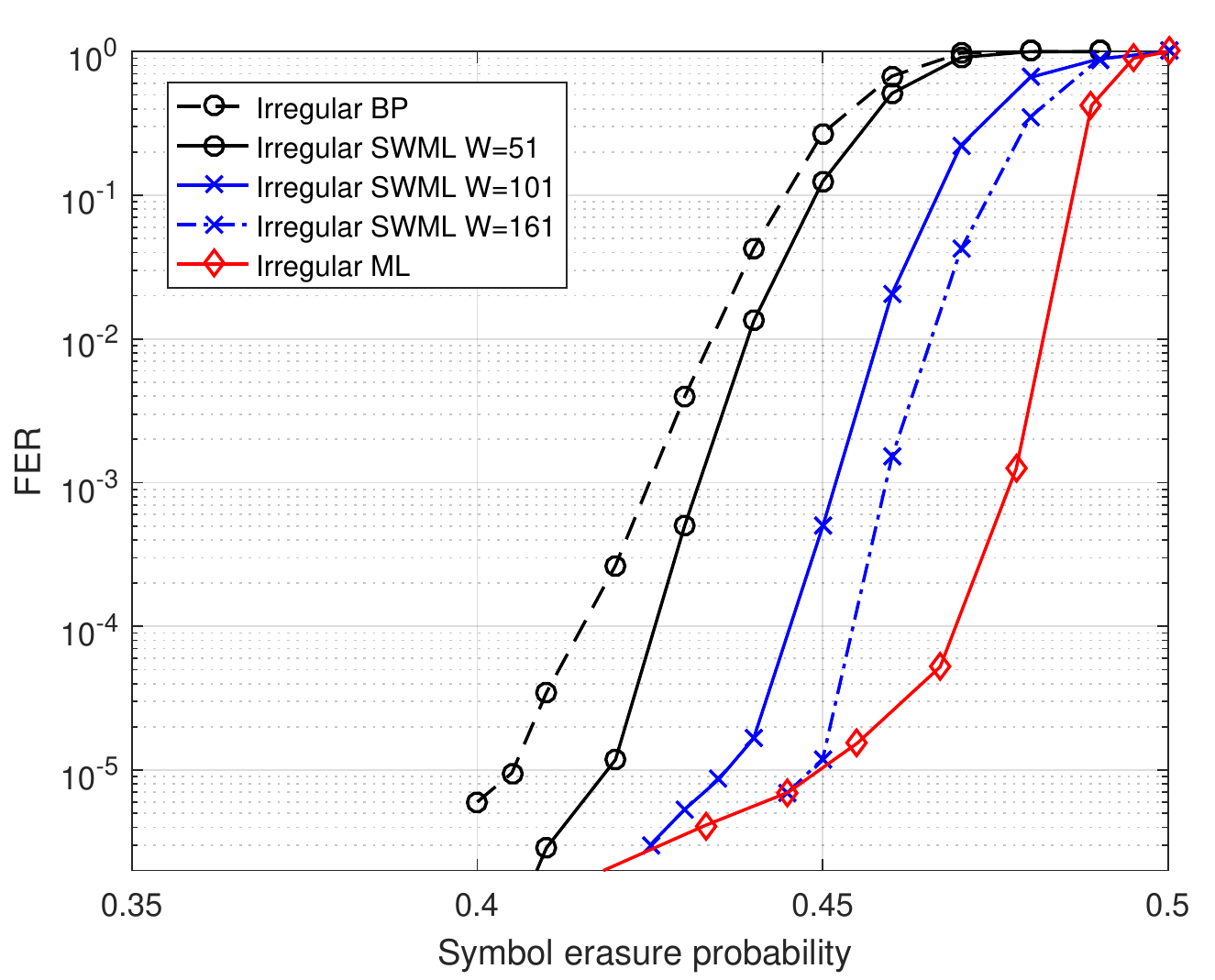}   
\caption{\label{SWML_ML} The simulated FER performance of  BP, ML, and 
SWML decoding  with different window sizes for an irregular $R=12/24$ QC LDPC code of length  $n=4800$.
  }
\end{center}
\end{figure}

\begin{figure}
\begin{center}
\includegraphics[width=100mm]{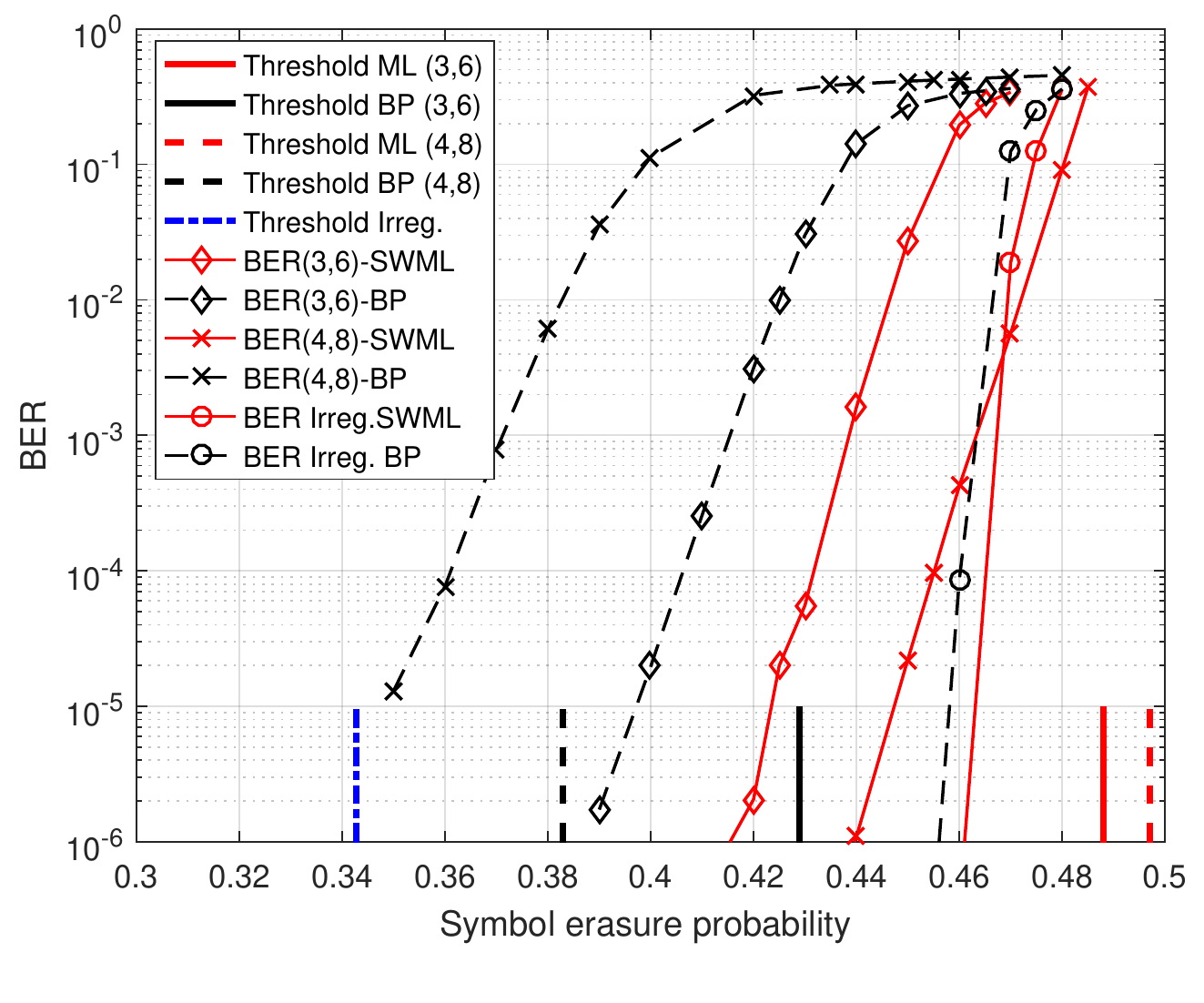}   
\caption{\label{c_th}  Comparison of  the simulated BER performance of  
SWML decoding of irregular QC LDPC codes of length  $n=60000-96000$ with 
BP and ML decoding thresholds.} 
\end{center}
\end{figure}

\begin{figure}
\begin{center}
\includegraphics[width=100mm]{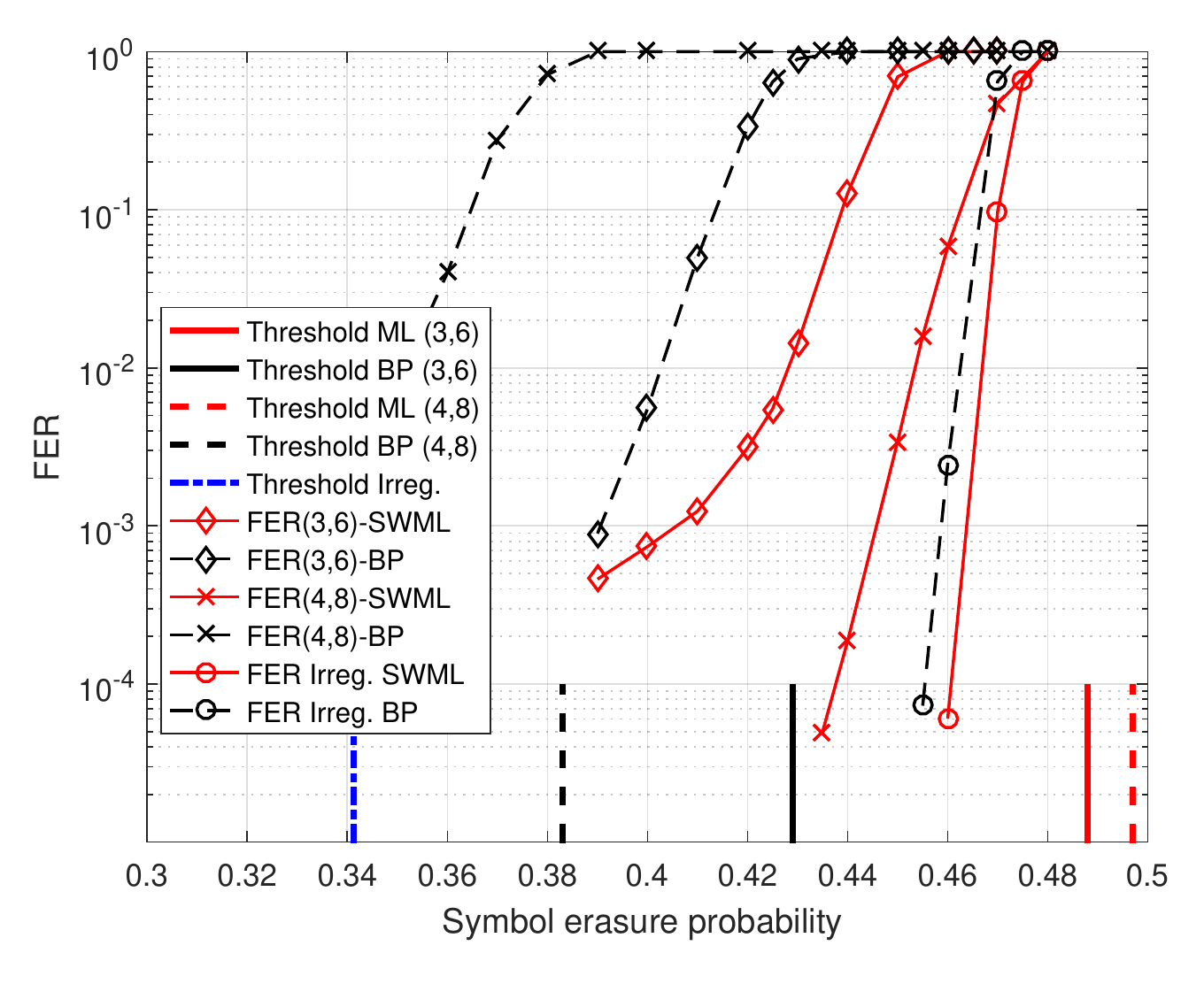}   
\caption{\label{c_th_fer}  Comparison of  the simulated FER performance of  
SWML decoding of irregular QC LDPC codes of length  $n=60000-96000$ with 
BP and ML decoding thresholds.} 
\end{center}
\end{figure}

\begin{figure}
\begin{center}
\includegraphics[width=100mm]{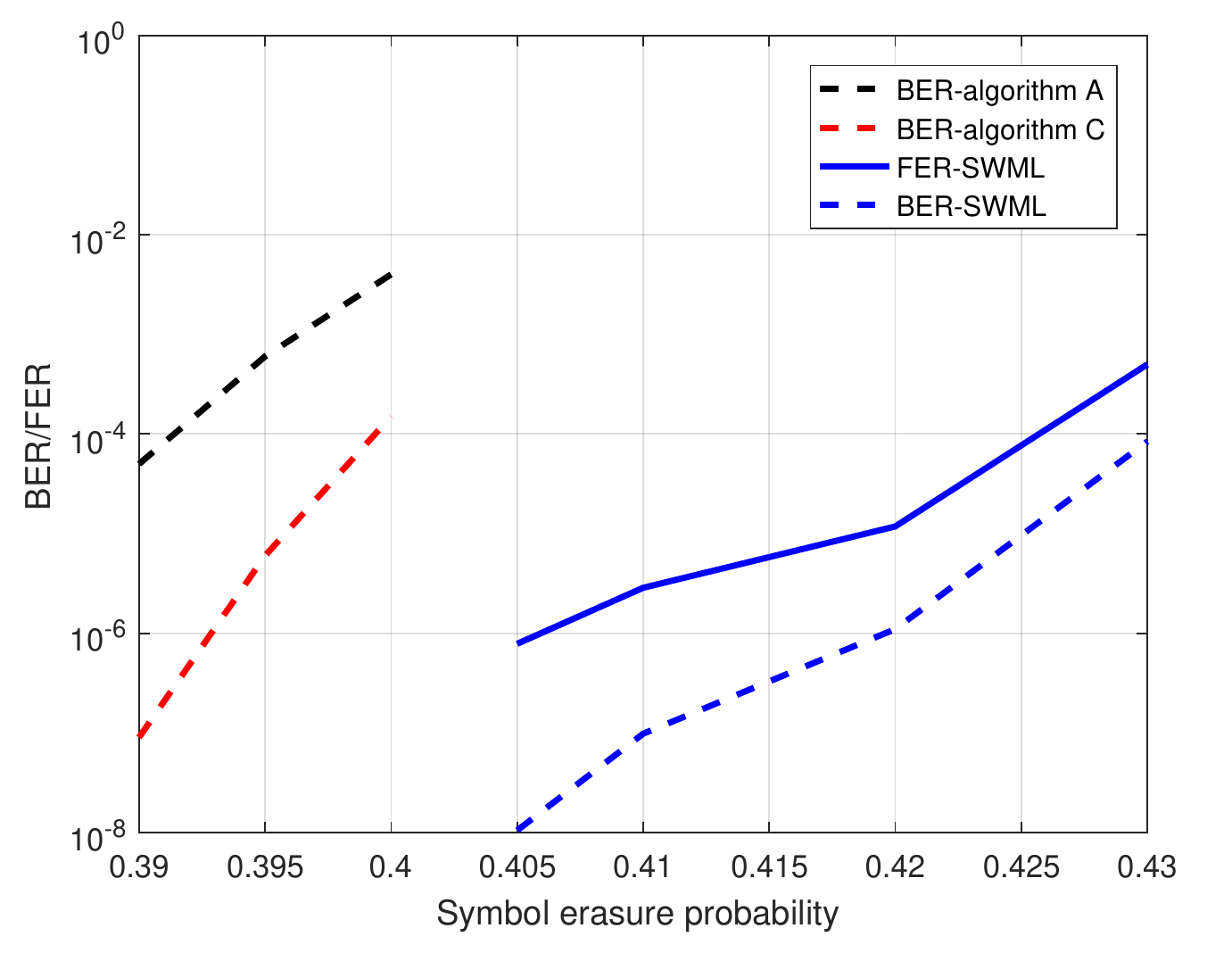}   
\caption{\label{comp_guess}  Comparison of  the simulated FER and BER performance of  
SWML decoding of the irregular QC LDPC code of length  $n=4800$ with the simulated BER performance of  
contraction-based decoding of random codes of length $10^{4}$.} 
\end{center}
\end{figure}

\begin{table}[h]
  \centering
  \caption{
\label{Table_WML}
Example parameters of SWML decoders for codes of  length $n=4800$.
}
\vspace{-2ex}
\begin{tabular}{|l|c|c|c|}
\hline
\T\B  Code and decoder  &\multicolumn{3}{c|}{\T\B  Codes} \\ \cline{2-4}
\T\B  parameters      & Irregular & DH   &  $(3,6)$-regular    \\  \hline
\T\B  Base matrix size $b\times c$& $12\times 24$ & $8\times 16$  &$3\times6$\\ \hline
\T\B  Syndrome memory $\mu$    &  $21$                 &   $15$   & 85\\ \hline
\T\B  Decoding window size $W$ & $51$               &   $81$  &  $171$ \\ \hline
\T\B  Window shift                   &  $24$               &     $16$ &    $6$ \\ \hline
\T\B  Overall TB length in bits $M_{0}\times c$ & $200 \times 24$   & $300 \times 16$  &$800 \times 6$  \\ \hline
\T\B  Maximum number of passes &    $15$  &  $15$ & 15 \\ \hline
\T\B  Minimum distance $d_{\min}$ &    $23$  &  $>24$ & 24 \\ \hline
\T\B Distance spectrum         &       $51, 306, \dots$        &          &     $2565,  0, 0, \dots$   \\ \hline
\T\B Stopping distance $d_{{\rm stop}}$ &    $19$  &  $>24$ & 24 \\ \hline
\T\B Stopping set spectrum                             &  $51, 0, 51, 0,$              &                    &  $2565,  0, 0, \dots$          \\   
       &       $255, 459, \dots$ &                      &             \\ \hline
 \end{tabular}
\end{table}

\begin{table}[h]
  \centering
  \caption{
\label{Table_WML_long}
Example parameters of SWML decoders for codes of  length $n=60000-96000$.
}
\vspace{-2ex}
\begin{tabular}{|l|c|c|c|}
\hline
\T\B  Code and decoder  &\multicolumn{3}{c|}{\T\B  Codes} \\ \cline{2-4}
\T\B  parameters      & Irregular & DH   &  $(3,6)$-regular    \\  \hline
\T\B  Base matrix size $b\times c$& $12\times 24$ & $8\times 16$  &$3\times6$\\ \hline
\T\B  Syndrome memory $\mu$    &  $98$                 &   $15$   & 85\\ \hline
\T\B  Decoding window size $W$ & $300$               &   $100$  &  $200$ \\ \hline
\T\B  Window shift                   &  $24$               &     $16$ &    $6$ \\ \hline
\T\B  Overall TB length in bits  $M_{0}\times c$& $4000\times 24$   & $ 6000\times 16$  &$10000 \times 6$  \\ \hline
\T\B  Maximum number of passes &    $15$  &  $15$ & 15 \\ \hline

\end{tabular}
\end{table}

 \section{Thresholds \label{Sec7}}

In order to  obtain an asymptotic upper bound on the error probability for $(J,K)$-regular
LDPC codes, we  use the following  inequality:
\begin{equation} \label{eq1}
\frac { \binom{n-\nu}{K}  }
                           { \binom{n}{K}}   
                            \le 
                            \left( \frac{n-\nu}{n}  \right)^K \; . 
\end{equation}

For the RU ensemble, it follows from (\ref{bound_LDPC}) 
{that}
\begin{eqnarray*}
P_e &\le& 
\sum_{\nu=1}^n \min\left\{1,B_\nu
 \right\}   \binom{n}{\nu} \varepsilon^\nu(1-\varepsilon)^{n-\nu}\\ 
&\le& 
{n \cdot \exp \left\{ -\min_{\nu} \left\{ \max  \left\{ T_{1},T_{1}-\log B_{\nu} \right\} \right\}  \right \} } \;,
\end{eqnarray*}
where
\begin{eqnarray*}
B_\nu&=& 2^{\nu-r} \left( 1+ \frac{\binom{n-\nu}{K}}{\binom{n}{K}} \right)^{r} \;, \\
T_{1}&=& -\log \binom{n}{\nu}-\nu \log \varepsilon -(n-\nu) \log (1-\varepsilon) \;.
\end{eqnarray*}

 
Denote by $\alpha=\nu/n$ the normalized number of erasures. The asymptotic error 
probability exponent can be written as
\begin{eqnarray}
E(\varepsilon)&=& \lim_{n\to \infty}\left\{ -\frac{\log P_e}{n}\right\} \nonumber \\
 &\ge&\min_{\alpha\in [0,1]}\left\{ \max \left\{F_{1}(\alpha,\varepsilon) \;,
 F_{2}(\alpha,\varepsilon) \right\} \right\}, \label{fer_exp}
\end{eqnarray}
where 
\begin{eqnarray}
F_{1}(\alpha,\varepsilon)&=&-h(\alpha)-\alpha \log \varepsilon -(1-\alpha)\log (1-\varepsilon) \;,\\
F_{2}(\alpha,\varepsilon)&=&F_{1}(\alpha,\varepsilon)-F_{3}(\alpha) \;,\\
F_{3}(\alpha)&=&\left(\alpha-\frac{J}{K}\right)\log2+\frac{J}{K} \left[  \log\left( 1+(1-\alpha)^K \right)\right] \;,\\ 
h(\alpha) &=&-\alpha \log \alpha - (1-\alpha) \log (1-\alpha) \label{fer_exp1} \;.
\end{eqnarray} 
In (\ref{fer_exp})--(\ref{fer_exp1})  all logarithms are to the base of $e$. 
The asymptotic decoding threshold is defined  as the maximum $\varepsilon$ 
providing $E(\varepsilon)>0$, or 
{as the} minimum $\varepsilon$ 
providing $E(\varepsilon)=0$.
It is easy to see that $F_{1}(\alpha, \varepsilon)$  is always positive except at the point  $\alpha=\varepsilon$ where $F_{1}(\alpha, \varepsilon)=0$, $F_{2}(\alpha,\varepsilon)>0$ for $\alpha<\varepsilon$, and $F_{2}(\alpha,\varepsilon)=0$ at  $\alpha=\varepsilon$  if  $F_{3}(\alpha)=F_{3}(\varepsilon)=0$.
%
%
In other words,  
a lower bound on the ML decoding threshold can be found as the unique solution of
the equation
\begin{equation} \label{bec_threshold}
\varepsilon=\frac{J}{K} \left[ 1-\frac{ \log\left( 1+(1-\varepsilon)^K \right)}{\log2} \right] \;.
\end{equation}
Notice that increasing  $K$ leads to the simple expression 
\[
\varepsilon  \xrightarrow[K\to \infty] {}\frac{J}{K} =1-R 
\]
for  the threshold, which corresponds to the capacity of the BEC. Numerical values for the lower bound from (\ref{bec_threshold}) on the ML decoding threshold for different code rates and different column weights are 
shown in Table~\ref{Table_BEC}.  
\begin{table*}[h] \small
\centering
\caption{
\label{Table_BEC}
Lower bounds from (\ref{bec_threshold}) on the ML decoding threshold for binary $(J,K)$-regular LDPC codes on the BEC. Exact values for the threshold computed from \cite[Lem.~4]{kud11} are given in bold in the parentheses.  Lower and upper bounds on the ML decoding threshold from  \cite[Table I]{sason2003parity} are reported in italics.}
\vspace{-2ex}
\begin{tabular}{|c|c|c|c|c|c|c|}
\hline
\multirow{2}{*}{
\T\B $R$ } &\multicolumn{6}{c|}{\T\B $J$} \\ \cline{2-7}
\T\B & $3$ & $4$ & $5$ & $6$ & $8$ & $9$ \\ \hline
\T\B $1/4$ & $0.74546793$ & & &$0.74998348 $ & &$0.74999993$ \\
\T\B & $\bf (0.74600970)$ & \bf--- & \bf--- & $\bf (0.74998850) $& \bf--- & $\bf (0.74999995)$\\ 
\T\B & $\it (0.744-0.7469)$ & & & & & \\ \hline
\T\B $1/3$ & & $0.66531587$ &&$0.66661773 $ &$0.66666485 $ & \\
\T\B & \bf--- & $\bf (0.66565559)$  & \bf---  & $\bf (0.66663252)$ & $\bf (0.66666541)$ &\bf--- \\ 
\T\B & & $\it (0.665-0.6657)$ & & & & \\ \hline
\T\B $1/2$ &$0.48696550$ &$0.49705261 $ &$0.49928578 $ &$0.49982316 $ &$0.49998898 $ &$0.49999724 $ \\
\T\B & $\bf (0.48815088)$ & $\bf (0.49774086)$ & $\bf (0.49948579)$ & $\bf (0.49987571)$ & $\bf (0.49999235) $& $\bf (0.49999809)$\\ 
\T\B & $\it (0.483-0.4913)$ & & & & & \\ \hline
\T\B $2/3$ &$0.31827745 $ &$0.32937002 $ &$0.33220831$ &$0.33300515 $ &$0.33330473 $ &$0.33332486 $ \\ 
\T\B & $\bf (0.31965317)$ &  $\bf (0.33025003)$ & $\bf (0.33251324)$ & $\bf (0.33310092)$ & $\bf (0.33331344)$ & $\bf (0.33332745)$\\ \hline
\T\B $3/4$ &$0.23601407 $ &$0.24609298 $ &$0.24882167$ &$0.24963402 $ &$0.24996371 $ &$0.24998853 $ \\ 
\T\B & $\bf (0.23732490)$ & $\bf (0.24694480) $ & $\bf (0.24913655)$ & $\bf (0.24973987)$ & $\bf (0.24997473)$ & $\bf (0.24999203)$\\ \hline
\end{tabular}
\end{table*}

%

 Unlike in {\cite[Eq.~(37), Table 1] {sason2003parity}} and \cite{measson2008maxwell }  the new lower bounds on the ML decoding thresholds presented in Table  \ref{Table_BEC}  are derived from the new upper bound on the average block error probability  for  the RU code ensemble. The thresholds   in {\cite[Eq.~(37), Table 1] {sason2003parity}} and \cite{measson2008maxwell } were obtained from  the bounds on the bit error probability of the average code in the ensemble. This explains the fact that  the lower  bounds in Table \ref{Table_BEC}  marked by asterisk differ from the best of the  lower bounds (shown in parentheses)  in \cite{measson2008maxwell } and {\cite[Eq.~(37), Table 1] {sason2003parity}}.

\section{Conclusion} \label{sec:conclu}
Both a finite-length and an asymptotic analysis of ML decoding performance of LDPC codes  
on the BEC have been presented. 
The obtained bounds are very useful, since unlike other channel models, for the  BEC,  
ML decoding can be implemented  for rather long codes. Moreover,  
an efficient  sliding-window decoding algorithm which provides near-ML decoding of very long codes 
is developed. Comparisons of the presented bounds with empirical estimates of the average error probability over sets of
randomly constructed codes have shown that the new bounds are rather tight at rates close to the channel capacity even 
for short codes. For code length $n>1000$, the bounds are rather tight  for a wide range of parameters.  The new bounds lead to a  simple analytical lower bound on  the ML decoding threshold on the BEC
for regular LDPC codes.

\section*{Acknowledgment}
The authors would like to thank A. Severinson for producing the ML decoding curve in Fig.~\ref{SWML_ML}.

\section*{Appendix~A}\label{Gallager_appr}
There is a weakness in the 
{proof of Theorem~2.4 in \cite{gallager} by Gallager, analogous to the one in the 
derivations (\ref{pdw})--(\ref{bin_spec}) above.  Formula} (2.17)  in
\cite{gallager} and (\ref{bin_spec}) in this paper  state that
the 
average number of 
weight-$w$ binary sequences which simultaneously satisfy all parity checks in  $J$  strips is 
\begin{equation} \label{gal_spec}
{\rm E}\{A_{n,w}\}=\binom{n}{w} \left[\frac{N_{n,w} } {\binom{n}{w} }  \right]^J \, ,
\end{equation}
	where  $N_{n,w}$ is the number of weight-$w$ sequences satisfying the parity checks of the first strip $H_{1}$. 
This formula relies  on the assumption that parity checks of strips are independent. 
It is known that this assumption is incorrect because 
the strips of the parity-check matrix are always linearly dependent (the sum of the parity checks of any two strips is  the all-zero sequence) and, as a consequence, 
{the actual rate} of the corresponding $(J,K)$-regular codes is higher than $1-J/K$. 
%
 %
The fact that we intensively use the strip-independence hypothesis in our derivations
motivated us to study deeper the influence of the strip independence assumption both on the conclusions in \cite{gallager} and on the derivations done in this paper. 



In order to verify  how this assumption influences the precision of estimates, consider the  
following simple example.

\begin{example}
Consider a $(3,3)$-regular code with $M=2$. 
The first strip is 
\[
\begin{pmatrix}
1&1&1&0&0&0\\
0&0&0&1&1&1
\end{pmatrix} \;.
\] 

The other two strips are obtained by random permutations of the columns of this strip.
In total there exist $(6!)^2$ LDPC codes, but most of the codes are equivalent. 
By taking into account that the first row in each strip determines the second row, we obtain that the choice of each code is determined  by the choice of the third and fifth rows of  the  parity-check matrix. 
Thus, there are at most $\binom{6}{3}^2=400$ equiprobable classes of 
equivalent codes. We compute the average spectra over  codes with a certain code dimension
and  the average spectrum  over all codes. 
The obtained results are presented in Table \ref{Spectra}. 
  
\begin{table}[h]
  \centering
  \caption{
\label{Spectra}
Spectra of $(3,3)$-regular LDPC codes of length $6$.
}
\vspace{-2ex}
\begin{tabular}{|c|c|c|c|c|c|c|}
\hline
\T\B Dimension &  Number of codes & Average spectrum \\ \hline
 \T\B $2$            &  $288$                   &
$\begin{pmatrix}  1& 0& \frac{1}{2} & 0 & \frac{5}{2}&  0& 0 \end{pmatrix}$\\ \hline 
\T\B $3$            &  $108$                   & 
$\begin{pmatrix}  1& 0& 2 & 0 & 5&  0& 0 \end{pmatrix}$\\ \hline 
  \T\B $4$            &    $4$                     &  
$\begin{pmatrix}  1& 0& 6 & 0 & 9&  0& 0 \end{pmatrix}$\\ \hline 
    \multicolumn{3}{|c|}{\T\B  Average parameters over the ensemble } \\ \hline
\T\B $2.29$       &    ---                 &   
$\begin{pmatrix}  1& 0& \frac{24}{25} & 0 & \frac{81}{25}&  0& 0 \end{pmatrix}$\\ \hline 
\end{tabular}
\end{table}

{Notice that the lower bound on the code 
rate $R\ge 1-J/K=0$, but since there exist  
at least two rows that are linearly dependent 
on other rows, a tightened lower bound on the code rate is $R\ge 1-4/6= 1/3$.}
Let us compare these  empirical estimates with the Gallager bound. 
The generating function for the first strip is 
\[
g(s)=1+N_{6,2}s^{2}+N_{6,4}s^{4}=1+6s^2+9s^4 \;.
\] 

According to (\ref{gal_spec})
\begin{eqnarray*}
{\rm E}\{A_{6,2}\}&=&\binom{6}{2} \left(\frac{6}{\binom{6}{2}}  \right)^3=\frac{24}{25} \;,\\
{\rm E}\{A_{6,4}\}&=&\binom{6}{4} \left(\frac{9}{\binom{6}{4}}  \right)^3=\frac{81}{25}\;,
\end{eqnarray*} 
which matches with the empirical average over all codes presented in Table \ref{Spectra}.
\end{example} 
These computations lead us to the following conclusions:
\begin{itemize}
\item
In the ensemble of $(J,K)$-regular LDPC codes there are codes of different dimensions.  
The average spectra depend on the dimension of the code and differ 
from the  average spectrum over all codes of the ensemble.
\item
The average over all codes may coincide with the Gallager estimate,  but 
does not correspond to any particular linear code. Moreover, the 
estimated number of codewords (sum of all spectrum components)
is not necessarily equal to a power of $2$.    
\end{itemize}

Notice that if $M$ is large enough, then the influence of strip dependence on the precision of the 
obtained spectrum estimate is negligible. However, 
if $\nu \ll M$, that is, in the low $\varepsilon$ region, the assumption of strip independence  
should be used with caution.

\section*{Appendix~B} \label{bound_proof}
\subsection*{Proof of Theorem~\ref{RankLdpc}}
{Assume that the number of erasures is $\nu > 0$.}
The error probability of ML decoding over the BEC is estimated 
as the probability that $\nu$ columns of the random parity-check matrix $H$ from 
the RU ensemble  corresponding to the erased positions are linearly dependent,  $\nu \le r$.  

Let $H_I$ be the submatrix consisting of the columns determined by the set $I$ 
of the erased positions, $|I|=\nu$. 
We can write 
\begin{equation}
\Pr\left(\mbox{rank}(H_I)  <  \nu\right|\nu)
\le
\sum_{\bs x_I \neq \bs 0}  \Pr \left(\bs x_I H_{I}^{\rm T}=\bs 0 \right|\nu) \;.
 \label{add2a}
\end{equation}

Consider a random vector   $\bs s=\bs x_{I}H_{I}^{\rm T}$.  Denote by $\bs s_i^j$ the subvector  $(s_i,\dotsc,s_j)$ of the vector $\bs s$.
The probability of  $\bs s$  being the all-zero vector is   
\begin{equation}
p(\bs s=\bs 0|\nu)  =  p(s_1=0|\nu) \prod_{i=2}^r p(s_i=0|\bs s_1^{i-1}=\bs 0,\nu) \;.  \label{prod1} 
\end{equation}
{Next, we prove that} $p(s_{1}=0|\nu)\ge p(s_{i}=0|\bs s_{1}^{i-1}=\bs 0,\nu)$, $i=2,3\dotsc,r$.
We denote by $\nu_i$ the number of erasures in nonzero positions of the $i$-th parity check.  
For the choice of a random vector $\bs x$ and a random parity-check matrix  
from the RU ensemble
the probability of a zero syndrom component $s_i$  is 
\begin{equation} \label{snu}
p(s_i=0|\nu_i)=\left\{
\begin{array}{ll}
1, & \nu_i=0   \\
\frac{1}{2}, & \nu_i>0
\end{array}
\right.\;.
\end{equation} 

First, 
{we observe} that 
for all $i$
\begin{eqnarray}
p(s_i= 0|\nu)&=& p(s_{i}=0|\nu_{i}=0,\nu)p(\nu_i=0|\nu)+
p(s_{i}=0|\nu_{i}>0,\nu)p(\nu_i>0|\nu) \nonumber \\
&=&
{1 \cdot p(\nu_i=0|\nu)+\frac{1}{2} \cdot (1-p(\nu_i=0|\nu))} \nonumber \\
&=&\frac{1+p(\nu_i=0|\nu)}{2}\;.      \label{prs1}
%
\end{eqnarray}
For all  $i\neq j$, let $K'$ denote the number 
{of row positions in which the corresponding elements 
either in row $j$ or in row $i$ of $H$ are nonzero.} Since $K' \ge K$ the following inequality holds:
\begin{eqnarray}
p(\nu_j=0|\nu_i=0,\nu)&=&\frac{\binom{n-K'}{\nu}}{\binom{n}{\nu}} \nonumber \\
&\le&\frac{\binom{n-K}{\nu}}{\binom{n}{\nu}}                             \nonumber \\
&=&p(\nu_j=0|\nu) \;.       \label{nuinuj}
\end{eqnarray}
%
%
%
For the second parity check, by using arguments similar to those in  (\ref{prs1}), we obtain
\begin{equation}
p(s_2=0|s_1=0,\nu)
=\frac{1}{2}\left(1+p(\nu_{2}=0|s_1=0,\nu)\right)  \;.  \label{secparcheck}
\end{equation}
The conditional probability in the RHS can be estimated as
\begin{eqnarray}\label{cond_prob}
p(\nu_{2}=0|s_1=0,\nu) &=& \sum_{\nu_1=0}^{\min\{K,\nu\}} p(\nu_{2}=0|s_{1}=0,\nu_1,\nu)p(\nu_1|s_1=0,\nu) \nonumber  \\
  &\stackrel{(a)}= & \sum_{\nu_1=0}^K
  p(\nu_{2}=0|\nu_1,\nu) 
  \frac{p(s_1=0|\nu_1,\nu) p(\nu_1|\nu)}  {p(s_1=0|\nu)} \;,
\end{eqnarray}
where equality (a) follows from the fact that
$p(\nu_{2}=0|s_{1}=0,\nu_1,\nu)=p(\nu_{2}=0|\nu_1,\nu)$. By substituting (\ref{snu}) into (\ref{cond_prob}),  we get
\begin{eqnarray} \nonumber
&&p(\nu_{2}=0|s_1=0,\nu)  \\
&=&\frac{p(\nu_2=0|\nu_1=0,\nu)p(\nu_1=0 |\nu)}{p(s_{1}=0|\nu)}+
\frac{\sum_{\nu_1=1}^{\min\{K,\nu\}} 
p(\nu_2=0|\nu_1,\nu) p(\nu_1|\nu ) }{2p(s_1=0|\nu)}\nonumber
\\
&=&\frac{p(\nu_2=0|\nu_1=0,\nu)p(\nu_1=0|\nu )+\sum_{\nu_1=0}^{\min\{K,\nu\}}
 p(\nu_2=0|\nu_1,\nu) p(\nu_1|\nu ) }
{2p(s_1=0|\nu)} \nonumber \;.
\end{eqnarray}
The second term in the nominator is equal to $p(\nu_2=0|\nu)$, and 
$p(\nu_i=0|\nu)$ does not depend on $i$. Thus, we obtain
\begin{eqnarray}
p(\nu_{2}=0|s_1=0,\nu) &=& p(\nu_2=0|\nu )
\frac{p(\nu_2=0|\nu_1=0,\nu)+1}{2p(s_1=0|\nu)} \nonumber  \\
&\stackrel{(a)}\le&
 p(\nu_2=0|\nu )
\frac{p(\nu_2=0|\nu)+1}{2p(s_1=0|\nu)} \nonumber  \\
&\stackrel{(b)}=&
 p(\nu_2=0|\nu ) \;,
\label{eq:prob_nu_2}
\end{eqnarray}
where inequality (a) follows from (\ref{nuinuj}) and equality  (b) follows from (\ref{prs1}).
{From~(\ref{prs1}), (\ref{secparcheck}), and (\ref{eq:prob_nu_2})} we conclude that
\[
p(s_2=0|s_1=0,\nu) \le p(s_2=0|\nu) \;. 
\]
Consecutively applying these derivations  for $i=3,4,\dotsc,r$ we can prove that
\[
p(s_{i}=0|\bs s_1^{i-1}=\bs 0,\nu) \le  p(s_{i}=0|\nu) \;,
\]
and then  from (\ref{prod1}) 
{it follows that}
\[
p(\bs s=\bs 0|\nu)
\le p(s_1=0|\nu)^{r} \, . 
\]
The  probability that the $i$-th row in $H_{I}$ has only zeros can be bounded from above by
\begin{equation*} 
p(\nu_i=0|\nu)=\frac { \binom{n-\nu}{K}  }
                           { \binom{n}{K}} \;,
\end{equation*}
and the probability that the entire sequence of length $\nu$ is a codeword (all $r$ components of the syndrome vector 
are equal to zero) is 
\begin{equation}
\label{zer}
p(\bs s=\bs 0|\nu)\le
2^{-r} \left( 1+ \frac{\binom{n-\nu}{K}}{\binom{n}{K}} \right)^{r} \; .
\end{equation}
%
{By substituting (\ref{zer}) into (\ref{add2a}),} we obtain 
\[
P_{e|\nu}=\Pr\left(\mbox{rank}(H_I)  <  \nu\right|\nu) \le 2^{\nu-r} \left( 1+ \frac{\binom{n-\nu}{K}}{\binom{n}{K}} \right)^{r}
\; ,
\]
{and the statement of Theorem~\ref{RankLdpc} follows from (\ref{genbound}).}

\section*{Appendix~C} \label{bound_proof_g}
\subsection*{Proof of Theorem~\ref{RankGal}}
Assume that 
{the number of erasures is $\nu>0$.} 
Let $H_I$ be the submatrix consisting of the columns numbered by the set $I$ 
of the erased positions, $|I|=\nu$. 
In Section~\ref{sec_bound} it is shown that the problem of estimating the FER of ML decoding
can be reduced to the problem of estimating the rank of the submatrix $H_I$. 
Let $H_{j,I}$ denote the $j$-th 
strip of $H_I$, $j=1,2,\dotsc,J$.
Denote by $\mu_i$ the number of all-zero rows in $H_{i,I}$, and define $\bs \mu=(\mu_1,\dotsc,\mu_J)$.

{Assume that the vector $\bs x$ is chosen uniformly at random
from the set of binary vectors of length $n$,  
and let $\bs x_I$ be the subvector of  $\bs x$ consisting of the elements numbered by the set  $I$.
Then, 
\begin{equation}
\Pr\left(\mbox{rank}(H_I)  <  \nu|\nu\right) 
\; \le \; 
\sum_{\bs \mu}  \Pr \left(\bs x_I: \bs x_I H_{j,I}^{\rm T}=\bs 0 \mbox{ for all } j \text{ and }\bs x_I\neq  \bs 0
  |\nu, \bs \mu
\right)  p(\bs \mu|\nu)   \; .\label{C1} 
\end{equation}
}

For the Gallager 
{ensemble} the conditional probability for the vector $\bs \mu$  given 
{that the
number of erasures is $\nu$ is} 
\begin{eqnarray*}
p(\bs \mu|\nu)&=&\prod_{i=1}^J p(\mu_i|\nu) \nonumber \\
&=&\prod_{i=1}^J  \binom{M}{\mu_i}\frac {\binom{n-\mu_i K}{\nu}}{\binom{n}{\nu}} 
=\prod_{i=1}^J  \binom{M}{\mu_i}\frac {\binom{n-\nu }{\mu_i K}}{\binom{n}{\mu_i K}}
\; ,
\label{C1a}
\end{eqnarray*}  
where we 
{take} into account that 
{the} strips are obtained by independent random permutations.
By using the inequality in (\ref{eq1}), we can bound this distribution from above as
\begin{eqnarray}
p(\bs \mu|\nu)&\le&\left[\prod_{i=1}^J \binom{M}{\mu_i}\right] 
\left[\prod_{i=1}^J \left( \frac {n-\nu}{n}    \right)^{\mu_i K} \right]
\label{C1b} \\
&\le&
\binom{M}{\mu/J}^J \left( \frac {n-\nu}{n}    \right)^{\sum_{i=1}^J \mu_i K} =
\binom{M}{\mu/J}^J \left( \frac {n-\nu}{n}    \right)^{\mu K}
\; ,
 \label{C1c}
\end{eqnarray}
where $\mu=\sum_{i=1}^J \mu_i$, and the second inequality follows from the fact that 
the maximum of the first product in (\ref{C1b}) is achieved when $\mu_1=\mu_2=\cdots=\mu_J=\mu/J$.

According to (\ref{snu})  each of 
{the} $M-\mu_i$ nonzero rows of 
{the $i$-th strip produces} a zero syndrome component
with 
{probability~$\frac{1}{2}$}. For a given $\boldsymbol{\mu}$, where $\sum_{i=1}^J \mu_i=\mu$,
$0\le\mu\le r$, 
the probability of having a zero 
syndrome vector can be upperbounded using a union bound argument as
{
\begin{eqnarray*}
\Pr \left(\bs x_I: \bs x_I H_{j,I}^{\rm T}=\bs 0 \mbox{ for all } j \text{ and } \bs x_I\neq  \bs 0|\nu, \bs \mu  \right) 
 &\le& \min\left\{1,
 \sum_{\bs x_I \neq \bs 0}
 \Pr \left(\bs x_I H_{j,I}^{\rm T}=\bs 0 \mbox{ for all } j|\nu, \bs \mu  \right) 
 \right\}
 \nonumber\\
&\le& \min \left\{1,
(2^{\nu}-1)\prod_{j=1}^J 2^{-M+\mu_j} 
\right\}
 \nonumber\\
&\le& \min \left \{1, 
2^{\nu-MJ+\sum_{j=1}^J \mu_j}
\right\} 
  = \min \left\{1, 2^{\nu-r+\mu} \right\}.
\end{eqnarray*}
}
From (\ref{C1}) it follows that 
{
\begin{equation}
\Pr\left(\mbox{rank}(H_I)  <  \nu\right|\nu) 
\le\sum_{\mu=0}^{r} 
\min\left\{1, 2^{\nu+\mu-r}  \right\} \sum_{\bs \mu: \sum_{j=1}^{J}\mu_j=\mu } p(\bs \mu|\nu) \;. 
\label{C2} 
\end{equation}
}
The total number of  different $\bs \mu$ with a given sum $\mu$ is equal to 
$\binom{\mu+J-1}{J-1}$.  From (\ref{C1c}) we obtain
\begin{equation}
\sum_{\bs \mu: \sum_{j=1}^{J}\mu_j=\mu } p(\bs \mu|\nu) 
\le%
\binom{\mu+J-1}{J-1}
\binom{M}{\mu/J}^J
\left(\frac{n-\nu}{n}\right)^{\mu K} \;.
\label{C4}
\end{equation}

Next, we use (\ref{genbound}), where for the conditional probability $P_{e|\nu}=\Pr\left(\mbox{rank}(H_I)  <  \nu|\nu\right)$
we apply estimates (\ref{C2}) and (\ref{C4}).
{
Each of the $\nu$ erasures belongs to $J$ rows, and  the total number $J\nu$ of nonzero elements 
are located in at least
$J\nu/K$ rows. Thus, the number of zero rows never exceeds $r-J\nu/K=J(n-\nu)/K$, which  explains the 
upper summation limit  in the second sum of (\ref{bound_Galn}).    
Thereby, we prove (\ref{bound_Galn})  of Theorem \ref{RankGal}.}

\bibliographystyle{IEEEtran}
\bibliography{IEEEabrv,listdec_bec4}

\end{document}